\documentclass[aps,a4paper,floatfix,nofootinbib]{revtex4}

\usepackage[utf8]{inputenc}

\usepackage{graphicx,color}
\usepackage{amsmath,amssymb,amsfonts,amsthm,amscd,bm}
\usepackage{booktabs}
\usepackage{cancel}

\def\tilde{\widetilde}
\def\bar{\overline}
\def\hat{\widehat}
\def\*{\star}
\def\[{\left[}
\def\]{\right]}
\def\({\left(}      

\def\){\right)}

\def\frac#1#2{\dfrac{#1}{#2}}
\def\inv#1{\dfrac{1}{#1}}
\def\half{\tfrac{1}{2}}
\def\d{\partial}

\def\2pi{\hbox{$2\pi i$}}

\def\dsl{\raise.15ex\hbox{/}\kern-.57em\partial}
\def\Dsl{\,\raise.15ex\hbox{/}\mkern-.13.5mu D}
      \def\CF{{\cal F}}
\def\CG{{\cal G}}   \def\CH{{\cal H}}   
      \def\CL{{\cal L}}
   \def\CN{{\cal N}}   \def\CO{{\cal O}}
      \def\CR{{\cal R}}
\def\CS{{\cal S}}   \def\CT{{\cal T}}

\def\2pi{\hbox{$2\pi i$}}

\def\dsl{\raise.15ex\hbox{/}\kern-.57em\partial}
\def\Dsl{\,\raise.15ex\hbox{/}\mkern-.13.5mu D}
\font\numbers=cmss12
\font\upright=cmu10 scaled\magstep1
\def\stroke{\vrule height8pt width0.4pt depth-0.1pt}
\def\topfleck{\vrule height8pt width0.5pt depth-5.9pt}
\def\botfleck{\vrule height2pt width0.5pt depth0.1pt}
\def\Zmath{\vcenter{\hbox{\numbers\rlap{\rlap{Z}\kern
    0.8pt\topfleck}\kern 2.2pt
    \rlap Z\kern 6pt\botfleck\kern 1pt}}}
\def\Qmath{
    \vcenter{\hbox{\upright\rlap{\rlap{Q}\kern3.8pt\stroke}\phantom{Q}}}}
\def\Nmath{\vcenter{\hbox{\upright\rlap{I}\kern 1.7pt N}}}
\def\Cmath{\vcenter{\hbox{\upright\rlap{\rlap{C}\kern
                   3.8pt\stroke}\phantom{C}}}}
\def\Rmath{\vcenter{\hbox{\upright\rlap{I}\kern 1.7pt R}}}
\def\Z{\ifmmode\Zmath\else$\Zmath$\fi}
\def\Q{\ifmmode\Qmath\else$\Qmath$\fi}
\def\N{\ifmmode\Nmath\else$\Nmath$\fi}
\def\C{\ifmmode\Cmath\else$\Cmath$\fi}
\def\R{\ifmmode\Rmath\else$\Rmath$\fi}
\def\barray{\begin{eqnarray}}
\def\earray{\end{eqnarray}}
\def\beq{\begin{equation}}
\def\eeq{\end{equation}}

\def\n{\noindent}
\def\kvec{{\bf{k}}}

\def\AA{\leavevmode\setbox0=\hbox{h}
\dimen0=\ht0 \advance\dimen0 by-1ex\rlap{\raise.67\dimen0\hbox{\char'27}}A}

\makeatletter
\def\iddots{\mathinner{\mkern1mu\raise\p@
\vbox{\kern7\p@\hbox{.}}\mkern2mu
\raise4\p@\hbox{.}\mkern2mu\raise7\p@\hbox{.}\mkern1mu}}
\makeatother

\theoremstyle{plain}

\theoremstyle{remark}

\def\gcal{\mathfrak{g}}

\def\rhovac{{\rho_{\rm vac}}}

\def\kcut{{k_c}}

\def\Z{\mathbb{Z}}

\def\adot{\dot{a}}
\def\addot{\ddot{a}}

\def\MBlackHole{M_\bullet}
\def\RBlackHole{R_\bullet}

\def\gfrak{\mathfrak{g}}

\def\R{{{\rm R}_{\scriptscriptstyle \infty}}}
\def\M{{{\rm M}_{\scriptscriptstyle \infty}}}
\def\Rinfinity#1{{{\rm R}^#1_{\scriptscriptstyle \infty}}}
\def\Einf{{{\rm E}_{\scriptscriptstyle \infty}}}

\def\Tinf{{{\rm T}_{\scriptscriptstyle \infty}}}

\def\mzeron{m_{\rm z}} 

\def\HoCMB{H_{0 ; {\rm CMB}}}
\def\HoSupernova{H_{0 ; {\rm SN}}}

\def\bhat{\hat{b}}

\def\OmegaLambda{\Omega_\Lambda} 

\def\zmax{z_{\rm max}}
\def\amin{a_{\rm min}}

\def\LambdaCDM{\Lambda{\rm CDM}}

\def\Egamma{E_\lambda}

\def\tmin{t_{\rm min}} 

\def\erfi{{\rm erfi}}

\def\Tmax{T_{\rm max}}

\def\darkuniverse{{\bf dark-universe}}

\begin{document}

\title{Quantum Vacuum energy as  the origin of  Gravity} 
\author{
 Andr\'e  LeClair\footnote{andre.leclair@cornell.edu} 
}
\affiliation{Cornell University, Physics Department, Ithaca, NY 14850,  USA} 

\begin{abstract}

We explore the idea that quantum vacuum energy $\rhovac$,  as computed in flat Minkowski space,  is at the origin of Gravity.         
We formulate a gravitational version of the electromagnetic Casimir effect, 
and provide an  argument for how gravity can arise from $\rhovac$ by  showing  how Einstein's field equations emerge  in the form of Friedmann's equations.  
This leads to the idea that Newton's constant $G_N$ is environmental,  namely it depends on 
the total mass-energy of the universe $\M$ and its size $\R$,   with  $G_N = c^2 \R/2\M$.  
This leads to  a  new interpretation of the Gibbons-Hawking entropy of de Sitter space, and also the 
Bekenstein-Hawking entropy for black holes,  
wherein the  quantum information ``bits" are simply quantized massless particles at the horizon with 
wavelength $\lambda = 2 \pi \R$.      We assume a recently proposed and well-motivated 
formula for $\rhovac \propto  \mzeron^4/\gcal$,  where $\mzeron$ is the mass of the lightest particle,  and
  $\gcal$ is a marginally irrelevant coupling.    
This leads to an effective,  induced RG flow for Newton's constant $G_N$ as a function of an energy scale,  
which indicates that $G_N$ {\it decreases} at higher energies until it reaches a Landau pole at a minimal value of the cosmological scale factor $a(t) > \amin$,   thus avoiding the usual geometric curvature  singularity at 
$a=0$.   The solution to the scale factor  satisfies an interesting symmetry between the far past and far future due to $a(t) = a(-t + 2 \tmin)$,   where $a(\tmin) = \amin$.     We propose that this energy scale dependent $G_N$ can explain the Hubble tension and we thereby constrain the coupling constant $\gcal$ and  its renormalization group parameters.      For the $\LambdaCDM$ model we estimate $\amin \approx e^{-1/\bhat}$ where $\bhat \approx 0.02$ based on the Hubble tension data.   
Comparison with other data besides the Hubble tension is considered.

\end{abstract}

\maketitle
\tableofcontents

\section{Introduction}

Gravity remains the least understood of the fundamental forces,   in spite of the discoveries of General Relativity (GR) 
and Quantum Mechanics (QM) over 100 years ago,  this year being recognized as the 100-th anniversary of  a complete theory of QM due to Schr\"odinger and Heisenberg,   although  the birth of QM goes back to Planck's black body studies at finite temperature.  
        Classical singularities  at the center of black holes and at the time of the so-called Big Bang have
not yet been understood and resolved in classical GR.      Many physicists  expect that such singularities will eventually  be resolved by a theory  of 
quantum gravity,  but it's fair to say that a complete theory of quantum gravity that can address these fundamental issues  has not been  forthcoming.         String theory still offers some promise and is the most well-motivated since it provides a consistent scattering theory of gravitons and more.     Nevertheless,  for these reasons it is worthwhile to pursue new ideas on the origin or emergence of  Gravity itself,   especially if such ideas hint   that perhaps Gravity doesn't need to be  quantized in the first place,  or at the very least is not yet needed to explain currently observable phenomena.     Such a theory should be able to explain the value of Newton's constant $G_N$ itself.   

 Like Gravity,   the Vacuum $|0\rangle$ is omnipresent.   The Vacuum  ``knows" the physical properties,  such as mass,  quantum numbers, and interactions  of any particle excitations created above it, and is perhaps the only 
entity  that  exists everywhere in the Universe and for all times.  The existence of the Quantum Vacuum guarantees that the allowed particle excitations above it,  such as electron/positron pairs,  
are identical throughout the Universe,  because they already  exist {\it virtually}  in both a physical and mathematical sense.       Everything else we observe are quantum excited states of the Vacuum above the ground state it represents,   including people,   and are typically  unstable.     
   Nature does not abhor a Vacuum as Aristotle first stated,     rather it depends on it.\footnote{To paraphrase with translation a recognizable 
    philosopher, 
{\it If you look deeply enough into the Void, It eventually looks back at you.}} 
   So it is reasonable  to suppose that the Vacuum,  in particular its ground state energy density $\rhovac$,   is  at the origin of Gravity.  If the only way to measure the ground state energy $E_0$  is with Gravity,   then reversing the argument,  perhaps quantum vacuum energy is the origin of Gravity.\footnote{In a certain sense,  the old aether  is back as the Vacuum itself.}  
 Deceptively  simple as it may appear, to study quantum field theory (QFT)  at finite temperature $T$ with periodic {\it spatial} boundary conditions, one
necessarily needs to study quantum fields in curved euclidean space where the euclidean time is compactified to a circle with circumference $\hbar c/k_B T$.
 In one  spatial dimension 
 the euclidean spacetime  is a  torus,   and the problem of $d=1$ conformal QFT  on a torus is  completely solved,  even though it is a  non-trivial problem involving modular invariance, etc.    
 From this point of view,  temperature is a fundamental  link between geometry and QFT. 
 For the same reason that one does not necessarily have to quantize the heat bath to study finite temperature QFT,    perhaps one does not have to quantize Gravity.  
If the Vacuum and its energy density  is the origin of Gravity itself,   then since the Vacuum is intrinsically Quantum,   one ponders whether  quantizing gravity is
essentially redundant.    
   Henceforth,   we set  aside the deep question as to whether Gravity  ultimately needs to be quantized since,  as we will see,   a resolution of the singularity at the origin of the so-called Big Bang can be understood without  it.\footnote{We take  ``quantum gravity" to mean quantization of 
   the metric gravitational field itself,  which necessarily implies spin-2 massless gravitons.    Other arguments that attempt to prove that gravity must be quantized are not  based on gravitons, but rather on the principles of  superposition and unitarity in  QM.  
  These arguments attempt to prove   that a purely classical gravitational field leads 
   to inconsistencies with the rest of the quantized world.     In a somewhat fanciful  thought experiment put forward by Feynman \cite{Feynman},  imagine  a massive ball  that becomes entangled with an electron,  where the electron  is in a superposition of two states with different paths,  thus the ball becomes a superposition of states with different locations.
   This would imply the gravitational field of the ball is in a superposition,   and this could in principle be detected by interacting gravitationally with a second ball.   
   These arguments rely on unitarity of QM,  and whether the Universe is an open or closed system is debatable. 
  For instance,  in  the search for a de Sitter/CFT correspondence,   it has been proposed that the CFT should be non-unitary \cite{Strominger}.     Recently a potential loop-hole in Feynman's thought experiment has been presented \cite{Oppenheim}.}         
This does not preclude the existence of classical gravitational waves that have been measured,   as they can be viewed as classical perturbations of the Vacuum as an elastic medium.

For purposes of illustration,   consider a universe which only consists of a single harmonic oscillator in its ground state  with energy $E_0 = \hbar \omega_0/2$,  in addition to an imagined observer.       Since classical mechanics is unaffected by shifts of the potential energy by an arbitrary constant,   $E_0$ is not measurable in classical physics.     If all that exists is the harmonic 
oscillator in empty space,   then $E_0$ also cannot be measured in a {\it single}  measurement  based on QM,  but can in principle be measured if coupled for instance to photons with repeated measurements.          There are two known general ways to measure $E_0$.    First,  one can couple the harmonic oscillator to a 
generic heat bath at temperature $T$.     Then one can in principle infer $E_0$ from the high temperature behavior of the partition function or  its  logarithm which is the free energy.     It's important that this is based on the free energy and not its derivatives,   and thus it  is not invariant under  shifts of the potential energy,  and this avoids 
some fine-tuning issues.       The second manner to measure $E_0$  is to couple the oscillator to Gravity,   since gravity originates from all the matter-energy in the Universe and is sensitive to its ground state energy.   
The fact that the ground state energy can be measured either by coupling to a heat bath or gravity 
suggests parallels between thermodynamics and gravity that have already been recognized,  in particular for the entropy of black holes.      
 If  our imagined observer  of the single harmonic oscillator thinks about natural units,   they would be based on $\hbar$, $c$, and $E_0$ which are complete,  as they lead to a fundamental length,  time and mass.

Based on the above remarks,   the initial  perspective on Gravity taken in this article  is based on the idea that perhaps we can learn more about the  foundations of Gravity  by considering the cosmological far future,   in contrast to the far past where much less is understood,  in particular the origin of the 
Big Bang and possibly inflation.        The late Universe is dominated by  so-called Dark Energy,  thus one should first  attempt to understand the full implications of this in a Universe  where Dark Energy is its sole component.    Below we will refer to such a hypothetical universe as the \darkuniverse,  and  it  should be viewed as a skeleton of a model of the Universe where there are no excitations over its ground state.   
 By assumption,    we will  equate Dark Energy  with the quantum vacuum energy density $\rhovac$ computed in quantum field 
 theory (QFT)  in flat local Minkowski space,     defined as    
\beq
\label{rhovacDef}
\langle 0 | T_{\mu \nu} | 0 \rangle = \rhovac \, \eta_{\mu\nu} 
\eeq
with the convention $\{ \eta_{\mu\nu} \} = {\rm diag} \{1, -1, -1, -1\}$
where the Vacuum $|0\rangle$ is the ground state for {\it all}  quantized fields,  {\it excluding} gravity.    
The  gravitational implications of the quantum field theoretic properties of $\rhovac$ will be then explored. 
Thus this article essentially explores  the question  
``Can Quantum vacuum energy be the origin of Gravity?"
The ideas  presented here  are most closely related to Andrei Sakharov's idea on ``induced gravity" \cite{Sakharov,Visser},   
 however they differ in many important details.\footnote{The parallels  with Sakharov's induced gravity ideas were  only pointed out to us after the first version of this article was made available on arXiv.  Sakharov's original article is extremely short with very little equations and is difficult to penetrate.     A nice explanation can be found in 
 \cite{MisnerThornWheeler}  page 426,  where it is explained how Newton's constant $G_N$ is viewed as a kind of elastic modulus.}
  For instance our model does not require a UV cut-off in momentum space as Sakharov's theory does,  since this UV divergence is controlled in the QFT through renormalization, and  this renders  
 $\rhovac$   well-defined both physically and mathematically.       This article can thus be viewed as  applying modern ideas of the renormalization group 
 to Sakharov's idea to deal with the cut-off,   which were not formulated until after Sakhavov's work by Wilson \cite{KWilson},   and   the resulting   renormalization group flow for $\rhovac$.   The theoretical framework is  still semi-classical gravity  where $T_{\mu\nu}$  for Dark Energy  is  identified with its vacuum expectation value.  
   We are unaware of applications of Sakharov's ideas  to the Hubble tension in the literature,  nor its prediction of the  minimal scale factor $\amin$ described below.

\bigskip

The main assumptions in this article are rather conservative and  minimal,   and it's useful to itemize  them here in order to understand how they fit into the existing literature. 

\bigskip\bigskip
\n {\it Assumption 1.}  ~~ Dark Energy,  or equivalently  the Cosmological Constant,  is equated with quantum vacuum energy density in flat Minkowski space, namely
the vacuum expectation value of the energy-momentum tensor \eqref{rhovacDef}.       This quantity is finite and well-defined in any QFT and studied in some detail 
recently for 4 dimensional spacetime \cite{ALrhovac1,ALrhovac2}.    There we obtained the formula 
\beq
\label{rhovacg}
 \rhovac =\frac{3}{4}  \frac{c^5}{\hbar^3}  \frac{\mzeron^4}{\gfrak} 
 \eeq
 where $\mzeron$ is the physical  (renormalized) mass of the lightest  {\it massive} particle,    and 
 $\gcal$ is  a dimensionless   interaction coupling constant,  or a function thereof if higher order corrections in perturbation theory are included.    The above quantity refers to the vacuum energy density for 
 {\it all}  known particles,    and the fact that the lightest mass particle $\mzeron$ appears is based on bootstrap principles.    This thus assumes there is a {\it single} particle mass scale that determines all other particle masses.     This is actually the situation in  the Standard Model of particle physics where all masses are thought to arise from the single scale through  the Higgs mechanism.    This assumption is even more correct  if there is a Grand Unified Theory of particle physics.    
    If the Vacuum and its energy density $\rhovac$  were  based on two or more completely independent and decoupled QFT's with different fundamental mass scales,   then the 
 formula \eqref{rhovacg} doesn't obviously apply since it becomes undetermined what $\mzeron$ actually is.   Thus we have to assume all particles can interact with each other,   otherwise one is dealing with separate,  distinct universes.     
 The formula \eqref{rhovacg} is not  ad hoc,   and was  well-motivated in 
  \cite{ALrhovac1,ALrhovac2}.     
 Since this formula is central to our study,  its underpinnings are reviewed and explained below in Section IVA.    It is not necessary for our purposes at this preliminary 
 stage to 
 definitively identify the $\mzeron$-particle as a known particle,   however we can constrain its mass,  equation  \eqref{mzOfg},  and it's coupling constants 
 based on Hubble tension data,  and we will have more to say about its properties below.

\bigskip
\n {\it Assumption 2.}  ~~ The coupling $\gcal$ is a marginally irrelevant coupling with 1-loop  renormalization group (RG) beta function:
\beq
\label{betagcal0}
\mu \d_\mu \gcal =  \frac{b}{2 \pi} \gcal^2 
\eeq
for some {\it positive} constant coefficient $b$,  
where increasing $\mu$ corresponds to a flow to higher energies.     Integrating the RG flow leads to the parameter $\bhat = b \,\gcal_0/2\pi$ where 
$\gcal_0$ is the value of $\gcal$ at the relevant energy scale {\it today}.   The parameter $\bhat$ is what modifies the Friedmann equations in the way presented below. 
Based on the Hubble tension we will constrain  $\bhat \approx 0.02$ which is quite small,   and justifies a 1-loop approximation.

\bigskip
\n {\it Assumption 3.}  ~~  There exists an energy scale of the Universe corresponding to a time dependent  temperature $T(t)$,   where as usual 
\beq
\label{Tdef0}
T(t) =  \frac{T_0}{a(t)} =  T_0 \, (1 + z(t) )
\eeq
where $a(t)$ is the scale factor and $z$ the redshift,  and $T_0$ is the temperature today.    As we will see,   de Sitter space has a natural temperature 
\eqref{HawkingT}.  
We assume $T_0$ is  the temperature of the Cosmic Microwave Background  (CMB) today,  $T_0 \approx 2.7$K,  however some predictions are not so sensitive
to the exact value of $T_0$ and it could in principle be the temperature based on neutrinos which is not so different.       
We mainly focus on the $\LambdaCDM$ model,   but we will assume that $T(t)$ is meaningful at energy scales above that of the CMB,  namely $z> 1100$,  at least up to some limit $\Tmax \approx T_0 (1 + \zmax)$ where $\zmax \sim  e^{1/\bhat}$.   (See below for explanations of $\zmax$.)

\bigskip
\n {\it Assumption 4.}  ~~ Matter and radiation are treated as particle excitations of the Vacuum above its ground state,  i.e. the Vacuum,   with energy density  $\rhovac$.      
We assume  $\rhovac$ is the vacuum energy density for {\it all} quantum fields,  and  all of the multiple  interactions and parameters  of the Standard Model of particle physics are implicit in $\rhovac$.  
 This allows for 
various deviations of \eqref{Tdef0}  at specific times where various phase transitions can occur,   such as the electro-weak phase transition and QCD phase transitions.    
We incorporate matter and radiation after first understanding a universe composed of only vacuum energy,  which we will refer to as  the \darkuniverse,  and  then subsequently add  matter and radiation  consistent with local energy-momentum conservation 
$\nabla^\mu T_{\mu\nu} = 0$,   which leads to the usual Friedmann equations with all forms of energy included.    As already stated,   the \darkuniverse~  should  be viewed as a skeleton for a complete universe which includes quantum matter and radiation.

\bigskip
\n {\it Assumption 5.}  ~~ We don't assume gravity is quantized,  in part simply because  we don't need it yet to resolve the curvature singularity of the Big Bang, as we will show.

\bigskip

\noindent       Under these assumptions all  of the QM involved is contained in the QFT computation of 
$\rhovac$ and it's renormalization group properties in Minkowski space.  We are thus  implicitly working in the framework of semi-classical gravity, and  simply work out the implications of these minimal,   well-founded assumptions,  to see where it leads us.          The parameters introduced are $\mzeron$,  which sets the scale of the cosmological constant,   $\bhat$,  and 
the  temperature $T$.    The Hubble constant $H_0$ is just an  initial condition for  the  Friedmann  differential equations at the present time $t_0$,  as is $T_0$.     
 Taking $\rhovac$ to be its currently measured value in cosmology,  namely 
 $ \rhovac = \Omega_\Lambda \rho_{\rm crit}$,  where $\Omega_\Lambda \approx 0.68$ one finds  
 \cite{Planck} 
 \beq
\label{mzOfg}
 \rhovac \approx  5.2 \times 10^{-10} \,   \frac{{\rm Joule}}{{\rm m}^3}, ~~~~~\Longrightarrow ~~\mzeron \approx  0.0024\, \gcal^{1/4}  \, {\rm eV}.
 \eeq 
 For $\gcal =\CO(1)$,   it is interesting to note that $\mzeron$ is on the order of proposed neutrino masses \cite{NeutrinoMasses},  however as previously stated, 
 the precise identification of the particle with mass $\mzeron$ will not be necessary for our purposes,  and is thus left as an open question. 
 Below we will propose additional constraints on the coupling $\gcal$.      
 From these parameters  alone one can develop a pseudo-realistic cosmos  which  can serve as a skeleton for our own Universe.   
 All of the ``microscopic" details of the evolution of the Universe obviously depend on the parameters,   interactions,  and all other bells and whistles of the complete QFT in Minkowski space of the Standard Model
 of particle physics.       As we will see,   within such a minimal  framework one can  at  least address some of the most profound  open problems of modern cosmology:    the origin of the Cosmological Constant,   the growing Hubble tension,   the resolution of $a(t)=0$ singularities associated with the so-called Big Bang,   what actually occurred before the Big Bang,   why the Universe is flat,  and whether current models of  inflation play a necessary role or what replaces it.        This framework may not provide final answers to all of  these deep questions,  but at the very least it is a novel approach to attempting to address them,   and not without some new testable predictions that we will present below.         

 Newton's constant $G_N$ is not on the list of basic parameters $\mzeron,  \bhat,$  and $T$ we just presented in the above Assumptions.
   In our proposed  framework $G_N$  it is not a 
 fundamental  {\it constant} of Nature,  and consequently nor is the Planck length  $\ell_p   =   \sqrt{\hbar G_N /c^3}$.    
From this perspective,  natural units should  be  based on $c, \hbar, {\rm and} ~ \mzeron$ rather than $c, \hbar, {\rm and} ~ G_N$.\footnote{From these,  one can define a fundamental length
$\ell_\ell =  2 \pi \hbar/\mzeron c$.    Based on the formula \eqref{mzOfg},  if $\gcal = 1$,  then $\ell_\ell = 5.2$ micrometers,  which is huge  compared to  the Planck length $\ell_p$.   What is more fundamental are mass/energy units.    Again for $\gcal =1$,   $\mzeron\approx 0.0024\, {\rm eV} /c^2 \approx 3.6 \times 10^{-36}$
\, grams,    which should be compared with 
the Planck mass  $m_p = \sqrt{\hbar c/G_N} \approx 2 \times 10^{-5}$ grams.   Based on our current understanding of fundamental  particle physics, 
 such as neutrino masses, 
a fundamental mass of about $0.002 \, {\rm eV}/c^2$   appears to us  more natural than $m_p$ which is about the mass of a large amoeba or very fine grain of sand.}   
If $G_N$ is not a fundamental constant,   then what determines its measured value?  
The first observations of Gravity were of course at the surface of the Earth which is manifested as a constant acceleration $g$ for all matter  regardless of mass or specific material.       Newton later understood $g$ to be ``environmental" rather than fundamental,   namely 
$ g =  G_N M_E/R_E^2 $,  where $M_E, R_E$ are the mass and radius of the 
Earth.         In Newton's universal  theory of Gravity, 
$G_N$ is considered a new fundamental constant of Nature,   and this view was adopted by Einstein in his formulation of 
General Relativity.\footnote{Throughout this paper,   $ G_N = 6.67 \cdot 10^{-11} {\rm  m^3}/{\rm kg\, s^2 }$ is Newton's  constant as measured today.      We keep the fundamental constants $\hbar$ and $c$ explicit in order to make certain conceptual points and to provide some  numbers with units.  ``Natural" units based on the Planck length $\ell_p$  are  not so natural for this article.}    
In contrast,   below we will argue that $G_N$ is also environmental,  namely 
 \beq
 \label{GMach}
  G_N = \frac{c^4}{2} \frac{\R}{\Einf} = \frac{c^2}{2} \frac{\R}{\M} 
 \eeq
 where $\Einf = \M c^2 $ is the total energy in the Vacuum for the entire Universe and $\R$ is a measure of its size.  These quantities will be defined more precisely below.\footnote{The subscripts $_\infty$  refer to the far future.} 
 All of the QM involved is incorporated into $\rhovac$ and thus  $\M$,      since $\M c^2 = \,  \tfrac{4}{3} \pi \Rinfinity{3} \, \rhovac$,  and $\R$ could
 be viewed as a kind of infra-red cutoff.    We derive the result \eqref{GMach} by formulating a gravitational version of the electro-magnetic Casimir effect,  
 wherein the ``conducting plates"  in a hypothetical experimental setup correspond to the horizon of the Universe itself.   Such a gravitational Casimir effect is formulated 
 {\it without} assuming Einstein's field equations.     Based on this gravitational Casimir effect,  we obtain the usual Friedmann equations,  which by comparison with the standard ones based on GR  leads to the identification of $G_N$ in the above equation \eqref{GMach}.   As we explain below,  the above formula \eqref{GMach} leads to a novel  and simple reinterpretation of the Gibbons-Hawking entropy for de Sitter space  \cite{GibbonsHawking} since the latter assumes $G_N$ is a fundamental constant.     In this reinterpretation,    the quantized information ``bits" are simply  quantized massless particles like photons at the horizon with wavelength $\lambda = 2 \pi \R$.     Such a  reinterpretation also applies to the
 Bekenstein-Hawking entropy \cite{Bekenstein,Hawking}  of black holes since the above equation \eqref{GMach} is the correct relation between $R$ and $M$ for a black hole.

 \def\cy{c_Y} 
 
 It is  instructive  to compare the perspective on Gravity developed  in this article with some proposed theories of quantum gravity,   since this clearly delineates some important distinctions.     We take ``quantum gravity" to mean a theory where the space-time metric $g_{\mu\nu} (x)$ itself  is quantized which implies the existence of
massless  spin 2  gravitons.   Based on this delineation,  quantum gravity plays no role in the present article,  and in spite of this our model avoids the usual curvature singularity when the scale fact $a(t) =0$.        Although the definition of the Planck length $\ell_p$,  defined from $G_N$ as   
 $G_N  =   c^3 \, \ell_p^2 /\hbar$,  is a priori  just based on dimensional analysis,   one expects that 
 $\ell_p$  is a relevant scale when considering graviton-graviton scattering.  
   A significant result of string theory,  which continues to be its main motivation,    is that it provides a consistent  prescription for the computation of graviton scattering that is 
 finite \cite{StringTheory}.   
 Thus   gravity ``emerges"  from string theory,   in the sense that Newton's constant $G_N$ is determined by matching graviton scattering with low energy gravity.      This leads to a formula for $G_N$ which is completely determined by physics at UV energy scales:
 \beq
\label{Gstring}
G_N  \sim  \cy  \, g_s^2 \, \, \frac{c^3 \ell_s^2 }{\hbar} 
 \eeq 
  where $\ell_s$ is the string length scale,  $g_s$   the string coupling,   and $\cy$ is a compactification factor depending on the Calabi-Yau manifold of the compact 6 spatial dimensions.\footnote{The fundamental length scale in string theory is related to the string tension $=1/2\pi \alpha'$ in the Polyakov action,  where $\alpha'$ is referred to as the Regge slope,   and $\alpha' = \ell_s^2$.\cite{StringTheory}}      The string  length scale $\ell_s$ must be close in value to the Planck scale $\ell_p$ to match low energy graviton scattering.         Thus $G_N$ is not yet predicted in string theory  since $\cy$ is unknown in particular  since there are many possible such compactifications.      
   Let us also mention Verlinde's  interesting entropic theory of gravity \cite{Verlinde} wherein gravity arises when one considers entropic forces  combined with the holographic ideas.    Here also $G_N$ ultimately is of the form \eqref{Gstring} with $\ell_s  \sim \ell_p$,  thus here also $G_N$ is determined from UV properties.          Finally,   for  loop quantum gravity, 
$G_N$ is treated as a fundamental constant in the  Einstein-Hilbert action.    Comparison of these graviton based expressions for $G_N$ with the formula \eqref{GMach} clearly indicates the significant differences with the perspective on Gravity presented in this article, since the formulas could not be more different. 
    First of all,  based on \eqref{Gstring},   Newton's constant becomes infinitely strong as $\hbar \to 0$ if the string scale $\ell_s$  is kept fixed.       
         In our formula \eqref{GMach},   all of the QM is contained in $\M$ through $\rhovac$.      Putting together some formulas we will obtain below,  one can write 
    $G_N  =   \gcal_0 \hbar^3 /(2 \pi c \mzeron^4 \Rinfinity{2})$.
where $\gcal_0$ is value of $\gcal$ at low energies.     In contrast to \eqref{Gstring},   here  note that $G_N \to 0$ as $\hbar \to 0$ in the latter formula,   
  which fits nicely  into the idea that Gravity originates from Quantum Mechanics in the form of $\rhovac$.     
  
With the above introductory remarks,   let us summarize the remainder of this article as outlined in the Table of Contents.
In the next section we formulate a gravitational version of the electromagnetic Casimir effect, 
and provide a heuristic   argument for how gravity can arise from $\rhovac$.  In particular we show how Einstein's field equations emerge,  in this context in the form of Friedmann's equations where $G_N$ is given by \eqref{GMach}.    
The formula \eqref{GMach} leads to new interpretation of the Gibbons-Hawking entropy of de Sitter space, 
wherein the  quantum informational ``bits" are identified as quantized massless particles like photons  at the horizon
with wavelength $\lambda = 2 \pi \R$,  as described in Section III.       
The analysis in Section II  is independent of the formula for  $\rhovac$ in \eqref{rhovacg},   and we turn to understanding the implications of it   in Section IV.   There we argue that the RG flow for the coupling $\gcal$  induces an effective  RG flow for Newton's constant $G_N$ as a function of an energy scale $\mu$.\footnote{We previously published some preliminary  work on the possible  implications of the RG flow of $\gcal$ for cosmology in  \cite{LeClairUniverse},   however the present article supersedes it.}    As is standard in Cosmology,   we tie this energy scale $\mu$  to the redshift 
$z$ and obtain equation  \eqref{CGdef2}.      In Section  V we solve exactly the modified Friedmann equations for the  \darkuniverse,   which  results in a significant modification of de Sitter space that is essentially an inverted  gaussian and at no time is $a(t) =0$,    which signifies no classical  type of geometric curvature  singularity since for all times 
$a(t)>\amin$.      
The solution to the scale factor satisfies an interesting symmetry between the far past and far future, 
namely $a(t) = a(-t + 2 \tmin)$ where $a(\tmin) = \amin$.    There is no Big Bang normally associated with the singularity $a(t) =0$,   rather at the time $\tmin$  the universe is at its 
hottest,   and is more  properly referred to as a hot Big Bang.   In our model since the solutions in past and future match at the time $\tmin$,  one can address what happened before the Big Bang.       Extending our solution to the far past,  the time evolution of 
our model universe is more analogous to the harmonic  swing of a pendulum.    Our interpretation of this is that the age of this  model universe is actually infinite since the solution to $a(t)$ extends to the far past without ever reaching the singularity at $a=0$.        In Section VI  we add matter and radiation,   
and due to the RG flow of $G_N$ in \eqref{CGdef2} there is again no $a=0$ singularity and the symmetry between the past and future continues to hold.      
This implies very small logarithmic corrections to the $\LambdaCDM$ model if $\bhat\equiv \, b \,\gcal_0/2\pi$ is small.              
We apply these ideas to the so-called Hubble tension,   wherein the discrepancy of measurements of
Hubble constant $H_0$ is proposed to arise  from the fact that the two contradictory measurements 
rely on data that refers to different epochs where Newton's constant is effectively different since the energy scales involved are different.    
In Section VII  we perform some checks of  our model by comparing with a variety of cosmological  and other types of experimental data.
The strongest support comes from recently observed trends in the Hubble tension for low-z supernovae.

\section{Emergence of Einstein's equations from a Gravitational Casimir effect}

In this section we formulate a gravitational version of the  electromagnetic Casimir effect.   These considerations are independent of the specific 
QFT details that lead to the formula \eqref{rhovacg}.       In this section  we only  assume that it is well-defined and finite,   and will turn to 
the implications of the formula \eqref{rhovacg} in the next section.    
 Without assuming Einstein's field equations,   we present heuristic arguments for how they  emerge in this context.    
 These arguments  lead to the formula \eqref{GMach}  for Newton's constant $G_N$ expressed in terms of length and energy scales in  
 our model universe.    The  discussion of this  section is not absolutely necessary for understanding the remainder of this article.    The reader may prefer to 
 simply assume Einstein's field equations and move onto the next section where QFT properties of $\rhovac$ are invoked.

Let us first review the electromagnetic Casimir effect,  which is well understood theoretically and has been measured \cite{Casimir}.
This  Casimir effect requires two uncharged,   parallel conducting plates in vacuum separated by a distance $L$.   The plates are required to radiate  and confine photons between 
the plates.      The energy density 
of quantized photons between the plates $\rho_{\rm rad}$ leads to a measurable force between the plates 
$F = \d_L E$ where $E = \rho_{\rm rad}  \cdot V$ where $V$ is the volume between the plates.     It is important to point out  that the theoretical predictions,  which are confirmed in the measurements,   do not measure the vacuum energy density $\rhovac$ but rather the energy density of {\it free}  photons to leading
order,   whose pressure is $p = w \rho_{\rm rad}$,  where $w=1/3$,  whereas genuine vacuum energy density has $w=-1$.       In fact,  since the  theory of  non-interacting photons is conformally invariant,    it's vacuum energy density 
$\rhovac =0$ in the leading conformal limit.

 For Gravity,   there is no well-formulated  and strict analog of the electromagnetic Casimir effect,  since gravitational waves and  hypothetical gravitons 
cannot be constrained by material boundaries like conducting plates.          In any case,  if any experiment  could be imagined that overcomes these 
obvious difficulties,   its effects would be much too small to be measurable if they originate from gravitons.        This leads  us to formulate a gravitational Casimir effect without material boundaries,  namely without plates,   that is sensitive to $\rhovac$,  where the latter is the vacuum energy density of {\it all} quantum matter and radiation excluding gravity.     Throughout this paper we assume $\rhovac>0$ such that we are dealing with de Sitter space.    Our working assumptions were presented in the Introduction.
 One additional assumption is that    
 the vacuum  has no time dependence originating  directly from  the time dependence of the resulting Einstein's equations,    namely   $\d_t \rhovac = 0$.   However as we will see in the next section,    tying the RG energy scale to the redshift $z$ leads to an additional induced time dependence.     
 
We carry out this construction  in arbitrary spatial dimension $d$ in order to make certain conceptual points.    Let us begin with only $1$ more spatial  dimension than the harmonic oscillator in QM discussed in the Introduction which corresponds to $d=0$.   
Thus,  consider  a relativistic QFT in $d=1$  spatial dimension,   from which one can compute $\rhovac$.  
 There are an infinite number of integrable QFT's  for $d=1$ where one can calculate $\rhovac$ exactly;  see for instance the examples in \cite{ALrhovac1,ALrhovac2} which are reviewed in Section IVA.        Let the spatial dimension be a finite size segment of length $2 R$.     Since total vacuum energy is $E =2 R \, \rhovac$,   there is a force $F = dE/dR= 2 \rhovac$ that can in principle be measured by moving the ``walls",  in this case simply the endpoints of the segment.   However if the endpoints do not consist of matter being held in place,  that is to say there is no analog of conducting plates,   
the effect of this force is expansion or contraction of the size of the segment,   depending on the sign of $\rhovac$.    This naturally leads us to introduce {\it geometry},  namely a  metric $g_{\mu\nu}$ which can incorporate this spatial expansion: 
\beq
\label{metric}
ds^2 = - g_{\mu\nu} (x) dx^\mu dx^\nu =  -c^2 dt^2 +  a(t)^2 dx^2 .
\eeq
The hypothetical experimental setup is the Universe itself,  since there are no conducting plates involved,  and as we will argue, 
 the cosmic horizon plays the role of such plates.    
The dimensionless scale factor  by definition satisfies    $a(t_0) =1$ at the present time $t_0$.    
  
 In order to make sense of Newton's 2nd law,    one needs to introduce an {\it inertial}  mass $M$.   One also needs to introduce a length scale $R$ such that $R \addot$ is an acceleration,  where as usual over-dots refer to time derivatives.   Equating the force $F$ with $2\rhovac$,  and requiring both sides scale the same way with the scale factor $a$, one obtains   
\beq
\label{Feq1}
MR  \,  \addot  = 2 a \,  \rhovac  ~~~ \Longrightarrow~~ \frac{\addot}{a} =  \frac{2 \rhovac}{MR} .
\eeq
We next  require 
$ \d_t \rhovac = 0 $.    One needs another equation to enforce this,   which is just $\addot/a = (\adot/a)^2$:  
\beq
\label{Feq3}
\( \frac{\adot}{a} \)^2  =  \frac{2 \rhovac}{MR}  .
\eeq
One can easily check that one time derivative of \eqref{Feq3} combined with \eqref{Feq1} implies $\d_t \rhovac =0$.

Let us extend the above arguments to $d$ spatial dimensions, and assume rotational symmetry,   where now $R$ is a radius.   Define $V_d (R)$ to be the volume of
a sphere in $d$ dimensions.   Then one has 
\beq
\label{d1} 
M R \, \addot = \frac{a}{d}\,   \frac{ \d  V_d (R)}{\d R} \rhovac   ,~~~~~ V_d (R) = \frac{2 \pi^{d/2} R^d}{d \,\Gamma(d/2)} ,
\eeq
where $\Gamma$ is the usual  $\Gamma$-function.  
The factor of $1/d$ on the right hand side of the first equation is due to the vectorial nature of the force.   
To summarize,   we have two equations after imposing $\d_t \rhovac =0$:  
\barray
\label{FriedmannD}
&~& \frac{\addot}{a} =    \frac{R^{d-2}}{M} \( \frac{2 \pi^{d/2}} {d\, \Gamma(d/2)} \) \rhovac
\\
&~& \(\frac{\adot}{a}\)^2  =  \frac{R^{d-2}}{M} \( \frac{2 \pi^{d/2}} {d\, \Gamma(d/2)} \) \rhovac .
\earray
Since $\rhovac$ is independent of time,   $H \equiv \adot/a $ is a constant.     If one identifies  $H=c/R$,    then the second equation just implies 
\beq
\label{rhovacConservation}
\rhovac  \cdot V_d (R) = Mc^2 ,   
\eeq
such that $M c^2$ can be interpreted as the total vacuum energy in a sphere of radius $R$.    

\def\xvec{\vec{x}}

Thus far,   Newton's constant $G_N$ has played no role,  since we have not assumed  Newtonian gravity nor Einstein's general relativity.       $G_N$ can be identified by comparing with Einstein's field equations for this particular metric.     
We adopt standard conventions for the Einstein-Hilbert action in $d$ spatial dimensions:  
\beq
\label{EinsteinHilbert}
\CS = \inv{16 \pi G_d} \int d^{d+1} x \,  \sqrt{-g} \, \( \CR + \CL_{\rm matter} \) .
\eeq
The metric is as in \eqref{metric},  with with $dx^2$ replaced with $d\xvec \cdot d \xvec$.   
Based on the above formulation of the gravitational Casimir effect,   there is no a priori reason to include a curvature term  in the spatial part of the metric proportional the standard $k\in \{-1, 0, 1\}$.   In other words $k=0$ is natural for our formulation,   since, from our perspective,   Gravity originates from $\rhovac$ in Minkowski space,  and the latter clearly 
has $k=0$.        
This leads to the field equations 
\beq
\label{EinsteinEq}
\CR_{\mu \nu} - \half g_{\mu\nu} \, \CR = \frac{ 8 \pi G_d}{c^{d+1}} \, T_{\mu\nu} .
\eeq
It's important to note that there is no room,  i.e. no a priori reason,  to introduce an additional classical  free parameter $\Lambda$ corresponding to an independent  cosmological constant term in Einstein's equations 
based on the above arguments.     Furthermore,  we already explained that once the QFT is decided upon,   there is no room for arbitrary shifts in $\rhovac$ since it
is based on the logarithm of its free energy density and not  its derivatives,  and this avoids the  fine-tuning issues normally associated with a variable free parameter 
$\Lambda$  corresponding to a  cosmological constant.  
If one carefully keeps track of the $d$-dependence,   this leads to the following Friedmann equations\footnote{The $d$ dependence of this formula should be a well-known result. 
One can find it for instance in \cite{GibbonsDdim}.} 
\barray
\label{Fried1d}
&~&\frac{\addot}{a} = - \frac{8 \pi G_d}{d(d-1) c^2} \[  (d-2) \rho + d \, p \]
\\
\label{Fried2d}
&~& \(\frac{\adot}{a}\)^2 =  \frac{16 \pi G_d}{d(d-1) c^2} \, \rho .
\earray
Specializing to vacuum energy with $T_{\mu\nu} = \langle 0| T_{\mu\nu} |0 \rangle$ in 
\eqref{rhovacDef},  where $p = - \rhovac$,    one obtains the equations \eqref{FriedmannD} 
with the identification 
\beq
\label{Gd}
G_d =  c^2 \frac{R^{d-2}}{M} \( \frac{(d-1) \pi^{(d-2)/2}}{8 \Gamma(d/2)} \) .
\eeq
 In 3 dimensions the above formula gives  $G_3 = c^2 R/2M$.   
Interestingly,   if $R,M$ are the radius and mass of a black hole,   then the above relation with $G_3 = G_N$ is their correct relation.    
We will return to this observation below where we reinterpret the entropy of  de Sitter space and black holes. 
      The case of $d=1$ is special since without matter the Einstein-Hilbert action is a topological invariant and there are no gravitational degrees of freedom.  This is reflected in $G_1 = 0$ in the above 
formula,   which is related to the fact that  $\adot/a \to 0$ as $R \to \infty$ in \eqref{Feq3}.\footnote{This basic fact  has led to alternative theories of gravity in one 
spatial dimension,   such as dilaton gravity,   wherein the $\CR$ term in the Einstein-Hilbert action 
is replaced by $\phi \CR$ where $\phi$ is a dynamical dilaton field,  such as in JT gravity 
\cite{JTgravity}.    One can show that equations 
\eqref{Feq1}, \eqref{Feq3} still apply in such a theory if one absorbs the $1/(d-1)$ into $G_1$.  In 4 spacetime dimensions there is no a priori reason to consider 
analogs of such models based on the ideas of this article,    such as Brans-Dicke gravity.}      

It remains to specify $R$ and $M$ in the above equations.      From our formulation of the gravitational Casimir effect,   it's clear that 
$M$ and $R$ refer to large scale properties of the universe,    and this  requires some input from the solution $a(t)$.        Recall that so far we have only considered vacuum energy density in
$T_{\mu\nu}$,  thus we are dealing with de Sitter space where  $H\equiv \adot/a$ is a constant by the second Friedmann equation  
\eqref{Fried2d}.     The  natural choice is then 
\beq
\label{RH}
R = R_H = \frac{c}{H}, 
\eeq   
which is referred as the Hubble radius.   
With the identification of $G_d$ in \eqref{Gd},   the equation \eqref{Fried2d}  is equivalent to  equation  \eqref{rhovacConservation}:
\beq
\label{Densityeqn}
\rhovac  V_d (R_H) = Mc^2 .
\eeq

In de Sitter space,  $R_H$ has several equivalent meanings which are all well-known and  we 
briefly review.    Here
\beq
\label{adS}
a(t) =  e^{H(t-t_0)}.
\eeq
$R_H$  equals the event horizon,    which marks the boundary beyond which an observer at a given time can never receive signals from events due to the accelerating expansion.    For a co-moving observer in de Sitter space,   the proper distance is given by 
\beq
\label{eventH}
d_{\rm event} = a(t) \int_t^\infty  \frac{c \,dt'}{a(t')}  = R_H .
\eeq  
The Hubble volume  $\tfrac{4}{3} \pi R_H^3$  
represents the volume that will remain causally connected to an observer as $t \to \infty$.   In other words,   it is the maximal volume of space that can influence an observer's world line.  
$R_H$ also  represents the distance at which objects recede in the expansion at the speed of light,  
which perhaps explains why the formula \eqref{GMach} is the correct relation for the mass
 $\MBlackHole$ and radius $\RBlackHole$ of a black hole if one identifies $\MBlackHole = \M,  
 \RBlackHole= \R$.      
Based on the above discussion we identify $R = R_H \equiv \R$,  and $M=\M$ defined through \eqref{Densityeqn} where the subscripts $_\infty$ refer to the far future.    
In 3 spatial dimensions,  this gives the formula in the Introduction:
\beq
\label{GMach2}
G_N =  \frac{c^2}{2} \frac{\R}{\M} .
\eeq

Let us for the moment express \eqref{GMach2}  as 
$G_N = 3 c^2 H^2/8\pi\rhovac$. 
If one relies only on experimental data in modern cosmology,   then the above expression \eqref{GMach2} for  $G_N$ is approximately correct with some assumptions,   however it should be realized that at this stage this  is a tautology since analysis of the cosmological data depends on $G_N$.      
First of all,  we have not yet included matter and radiation as excitations over the Vacuum,  where $H$ is no longer a constant, and 
we will return to this below.     Let us just  point out that if one takes the  experimental value \eqref{mzOfg} and identifies 
$H = \sqrt{\OmegaLambda} H_0$ where $H_0$ is Hubble constant as measured by the CMB \eqref{Hos} and
$\OmegaLambda = 0.68$,   
then the above expression evaluates to $G_N = 6.72 \times  10^{-11} $ ${\rm m}^3/{\rm kg}\,{\rm s}^2$.   Let us note that 
the same computation using $\HoSupernova$ in \eqref{Hos} yields a higher value for $G_N$.     This will be addressed  below in connection with the ``Hubble tension"  by taking into account the explicit expression \eqref{rhovacg} for $\rhovac$.

\section{A  reinterpretation of the  entropy of de Sitter space and black holes}

There are some important results in semi-classical gravity where the gravitational field is not quantized, 
namely  they do not rely on the existence of gravitons,  but the result still depends on the Planck length 
$\ell_p$,    such as  the Bekenstein-Hawking entropy and temperature for black holes \cite{Bekenstein,Hawking}.     In this present context,   
the analog is the Gibbons-Hawking entropy for de Sitter space \cite{GibbonsHawking}.    Since we have brought into scrutiny whether  $G_N$ and thus  Planck length $\ell_p$ 
are fundamental parameters,   this led us to re-interpret these entropies,   since they are expressed in terms of $\ell_p$.  
The expression \eqref{GMach} for $G_N$ leads to such a  new  interpretation that is very simple and 
and hints at an interpretation of the entropy in terms of bits of {\it quantized energies} rather than 
{\it quantized spacetime}.     We should emphasize that this reinterpretation is merely algebraic at this stage and is not the main thrust of this article,  though worth pointing out.    We do not attempt to derive 
the Gibbons-Hawking entropy from a fundamental counting of microstates $\Omega$ such that 
$S_{GH} = k_B \log \Omega$,   and we set this issue aside for the remainder of this article,  but will return to it in  a separate article under preparation.\footnote{In the published version of this article in JHEAP \cite{JHEAP},   we incorrectly used $\hbar$  rather than Planck's constant $h$ in the formula for 
$E_\lambda$,    which amounts to expressing the main result  for $S_{GH}$ in terms of the {\it reduced} wavelength $\lambda$.     Although the formulas in \cite{JHEAP} are correct as written,   the proper interpretation should be based on the actual wavelength not the reduced one,  and this is adhered to here.}

The Vacuum $|0\rangle$ itself has zero entropy since it is a single state.\footnote{
The bulk entropy is zero as can be seen from the relation $\CS = \beta^2 \d_\beta \CF = \beta ( \rho + p) = 0$ if $\rho = \rhovac = -p$,    where 
$\beta = 1/k_B T$,  and   $\CS$ and $\CF$ are the entropy and free energy densities.}    
However the spacetime  geometry does have a horizon $\R$,   and a non-zero entropy should be associated with it.   
Following Bekenstein's original reasoning based on the Shannon entropy $-\sum_ i  p_i \log p_i$ in information theory,    let us equate the  entropy $S$  to  the number of ``bits"   $N_{\rm bits}$
of information on the 
horizon: 
\beq
\label{Sbits}
S/k_B \sim   N_{\rm bits} .
\eeq
 The Gibbons-Hawking entropy  $S_{GH}$  was  proposed to be 
\beq
\label{SBH}
 S_{GH} \equiv  \frac{k_B}{4}  \frac{A}{\ell_p^2} = k_B \( \frac{\pi c^3}{\hbar} \) \frac{\Rinfinity2}{G_N}. 
 \eeq
 where $A = 4 \pi \Rinfinity2$.      Here each bit  is 
  commonly interpreted as a pixelation of the area $A$ into quantized units of area $4 \ell_p^2$,  suggesting that spacetime itself is somehow quantized into these minimal area units at the horizon.  The overall factor of $1/4$ is the hardest to explain in the above formula.        Such a quantization of spacetime itself remains a  somewhat vague notion,  and difficult to make sense of  
  mathematically.    A more appealing interpretation would correspond to quantized {\it energies} rather quantized spacetime.            

The fomula \eqref{GMach} leads to a very different interpretation of what the bits actually are for the
Gibbons-Hawking entropy \eqref{SBH}.     
    Let us identify such bits with massless modes,  or particles   like photons,   at the horizon.     
The smallest energy of  such a particle is $\Egamma = 2 \pi \hbar c/\lambda$ where the wavelength is quantized as $\lambda = 2 \pi \R$ since $2 \pi \R$ is the circumference.       The entropy should be extensive and thus proportional to $\Einf$.      
Based on \eqref{Sbits} we thus propose the entropy is given by the simple formula 
\beq
\label{MyEntropy}
S_{GH}/k_B  = 2 \pi 
 \,  \frac{\Einf}{\Egamma},       ~~~~~~{\rm with} ~ \Egamma =   \frac{2 \pi \hbar c}{\lambda} , ~~~~
\lambda = 2 \pi \R, 
\eeq
where as above $\Einf = \M c^2 $.   
Here,   the bits are indeed interpretted as quantized energies rather than hypothetical quantized patches of spacetime. 
Remarkably,   the above formula is identical to the Gibbons-Hawking entropy,  including  the overall $1/4$,
if one identifies $G_N$ as above \eqref{GMach}:
\beq
\label{MyEntropy2}
S_{GH} =  k_B  \frac{\M c^2 }{\Egamma} =  k_B \(\M c^2 \)  \( \frac{2 \pi \R}{\hbar c}\)  = 
   \frac{k_B}{4}  \frac{A}{\ell_p^2}, 
\eeq
where in the above equation $\ell_p = \sqrt{\hbar G_N/c^3}$  and $G_N$  is given by \eqref{GMach}. 
The analog of the Hawking temperature $T\equiv \Tinf$ as usual follows from  $d \Einf = \Tinf  d S$  which gives the simple result 
\beq
\label{HawkingT}
 k_B \Tinf =  \Egamma / 2 \pi  = \frac{\hbar c}{2 \pi \R} .
\eeq

\def\ttilde{\tilde{t}} 
The temperature $\Tinf$ is precisely the unique Matsubara temperature associated with de Sitter space. 
This is based on the topology of de Sitter space $\mathbb{R}   \times  S^3$ which has closed space-like curves,  in contrast to anti-de Sitter which has closed time-like curves.   Setting $\hbar = c = k_B = 1$, 
 this leads to $\beta \equiv 1/ \Tinf = 2 \pi \R$.     The $\beta$-periodicity is not manifest in the 
 standard cosmological coordinates based on the metric in \eqref{metric},    since it does not cover all 
 of de Sitter space.      One needs global coordinates for a universal cover.    A  known standard coordinate 
 transformation leads to 
 \beq
 \label{metric2}
 ds^2 = - dt^2 + R^2 \cosh^2 ( t/R ) \,  d\Omega_3^2 .
 \eeq
    Making the Wick rotation $t \to -i \tau$,    the $\cosh$ becomes $\cos (\tau/R)$ which has the proper periodicity $\tau \to \tau + \beta$ for a finite temperature Matsubara formalism with 
 $\beta = 2 \pi R$.   
 Since this temperature is fundamental,   one can effectively reverse the argument and obtain 
$S_{GH}$ from $\Tinf$.    
  Needless to say,  this interpretation of the bits in the Gibbons-Hawking entropy is much more tangible than the quantization of spacetime itself.   In a weak sense this is a sort of dS/CFT correspondence if the massless modes at the horizon correspond to a conformal field theory.

$\Tinf$  is an extremely low temperature, consistent with the fact that in the far future the temperature of the Universe is expected to  go to zero. 
For instance,   approximating $H= \sqrt{\OmegaLambda} H_0$ with $\OmegaLambda = 0.68$,
and $H_0$ determined from CMB data, see eq.  \eqref{Hos},   one finds $\Tinf \approx 2.2 \times 10^{-30}$\,K.     

In light of the above observations,   let us also attempt to reinterpret the Bekenstein-Hawking entropy of
black holes in a similar fashion.\footnote{For super massive black holes,   the quantization of spacetime itself into bits at the  super large horizon  seems to us as rather implausible.}     
    The reason this is possible is that \eqref{GMach} is the correct relation between the mass $\MBlackHole$ of the black hole and its radius $\RBlackHole$ if one makes the replacement 
$\M,\R \to \MBlackHole,\RBlackHole$:
\beq
\label{GBH}
G_N = \frac{c^2}{2} \frac{\RBlackHole}{\MBlackHole} .
\eeq
       From this perspective,   $G_N$ is a  constant which is the same for every black hole in the Universe.    Repeating the above arguments for  the 
Bekenstein-Hawking entropy,
\beq
\label{BekensteinHawkingS}
S_{BH} =  \frac{k_B}{4}  \frac{A}{\ell_p^2} =  k_B \( \frac{\pi c^3}{\hbar} \) \frac{\RBlackHole^2}{G_N} 
= 2 \pi \, k_B \,  \frac{E_\bullet}{\Egamma},       ~~~~~~{\rm with} ~ E_\bullet  = \MBlackHole c^2 ~~{\rm and} ~~
 \Egamma =  \frac{\hbar c}{\RBlackHole} ,
 \eeq
where we have used \eqref{GBH} in the second equation.    
However let us point out that the 
 Hawking temperature $T_H$ of the black hole  differs from equation  \eqref{HawkingT} by a factor 
 of $2$,   namely
$k_B T_H =  \hbar c/4 \pi  \RBlackHole$.   This can be attributed to the fact that due to the point-like singularity at the origin of the Schwarzschild metric,  the geometry is intrinsically different,  in particular the formula for the temperature based on surface gravity differs by a factor of $2$.   This factor of 2 can also be explained thermodynamically by writing the entropy as $S_{BH}  = 2\pi \MBlackHole \RBlackHole$  such that 
$dS = 2 \pi (\RBlackHole d\MBlackHole + \MBlackHole d \RBlackHole)$.  If one imposes 
$dG_N = 0$ with $G_N$ given in  \eqref{GBH},  then $\MBlackHole d \RBlackHole = \RBlackHole d \MBlackHole$,     which implies  $d E_\bullet  \neq T_{H}   dS_{BH}$ due to this factor of $2$.

\section{Induced renormalization group flow for Newton's constant}

\def\rhovaccutoff{\rho_{\rm vac,cutoff}}
\def\kcut{k_c}

\subsection{Justifying our  formula for $\rhovac$.}

In this subsection we review the analysis that led us to propose the formula \eqref{rhovacg} for $\rhovac$ in \cite{ALrhovac1,ALrhovac2}.    
As motivation,   let us  simply consider a real scalar field $\phi$ in $d+1$ spacetime dimensions with euclidean action
\beq
\label{Jrhovac1}
\CS = \int d^{d+1} x \, \( \half \d_\mu \phi \d_\mu \phi  + V(\phi) \).
\eeq
Suppose the potential is simply a mass term $V(\phi) = m \phi^2 /2$ such that spectrum consists of 
particles of energy $\omega_\kvec = \sqrt{\kvec^2 + m^2}$.    Treating the free quantized field as a collection of harmonic oscillators 
 with energy $\hbar \omega_\kvec/2$,  naively the vacuum energy density is  
\beq
\label{rhoWein1}
\rhovaccutoff = \frac{s}{2} \int_{|\kvec|  \leq \kcut}  \frac{d^d \kvec}{(2 \pi)^d} \, \sqrt{\kvec^2 + m^2 },
\eeq
where $s=+/-$ corresponds to bosons/fermions.  Since the above integral is divergent,  we have introduced a cut-off $\kcut$.
The leading terms for $\kcut \gg m$ are 
\beq
\label{rhoWein2}
\rhovaccutoff =  s
\begin{cases}
\frac{\kcut^2}{4 \pi} + \frac{m^2}{4 \pi} \,  \log (2 \kcut/m )   ~~~~~~~~~(d=1) \\ 
\\
\frac{\kcut^3}{12 \pi} - \frac{m^3}{12 \pi}  ~~~~~~~~~~~~~~~~~~~~~~~(d=2) \\
\\
\frac{\kcut^4}{16 \pi^2} - \frac{m^4}{32 \pi^2} \, \log ( 2 \kcut /m  ) ~~~~~(d=3)
\end{cases}
\eeq
The above calculation is what  led Weinberg to state  a Cosmological Constant Problem \cite{Weinberg}.  
Namely,  if $\kcut$ is taken to be the Planck scale,   then the leading term is off by $120$ orders of magnitude.   
There are two obvious problems with this calculation.    First of all,   there is no justification for $\kcut$ to be the Planck scale since
we are dealing with a free quantum field theory in Minkowski space with no gravity.     Secondly,   one is accustomed to ultra-violet divergences in 
QFT and how to deal with them in order to compute physical quantities.     In such a renormalization procedure the leading $\kcut^{d+1}$ is discarded.
However note that for $d+1$ even there is an unavoidable $\log \kcut$ divergence,   unlike $d+1$ odd,  and further renormalization is required in order to obtain
something finite and independent of the cut-off.   However if there are no additional interactions in $V(\phi)$,  there is no clear and physical way to remove 
the $\log \kcut$ divergence.   Based on the above equations \eqref{rhoWein2},   the issue of how to deal with the remaining $\log \kcut$ divergence is essentially identical in $d=1$ and $d=3$ spatial dimensions,
thus one can gain insights on the problem from interacting QFT's in $d=1$ that are exactly solvable,  i.e. integrable.

Firstly,  one  can argue that in $d+1$  spacetime dimensions $\rhovac \propto m^{d+1} /g$ where $m$ is a physical mass,   and $g$ is an interaction coupling,  
such that $\rhovac$ diverges as $g\to 0$ reflecting the unavoidable divergence in the free theory displayed in \eqref{rhoWein2}.      If $m$ is a physical mass, 
then by dimensional analysis,   $g$ is a dimensionless coupling for an interacting theory.   
That  $\rhovac \propto 1/g$ implies that it is intrinsically non-perturbative.   
It turns out this is exactly what occurs for integrable QFT in 2 spacetime dimensions \cite{ZamoZamo0,ZamoTBA}.     Namely for all these integrable theories, 
\beq
\label{rhovac2D}
\rhovac =  \frac{m_1^2}{2 \gcal}~~~~~~~~~~~~~~~~~~~~~~~ (d=1)
\eeq
where $m_1$ is the mass of the lightest  massive particle,  and $\gcal$ represents a dimensionless coupling,   which is a non-perturbative function of the couplings $g$ in
$V(\phi)$.     The formula \eqref{rhovac2D} can be derived from the thermodynamic Bethe ansatz,  which was first formulated by Yang and Yang \cite{YangYang}, 
and generalized to relativistic theories by Al. Zamolodchikov \cite{ZamoTBA}.    That  $m_1$ is the mass of the lightest {\it massive} particle  follows  from the fact that {\it any} particle can
probe the vacuum to determine $\rhovac$ and the result should be the same,    combined with the S-matrix bootstrap which implies that in principle the S-matrix for 
higher mass particles can be obtained by bootstrapping the lightest mass particle.   For integrable theories all of this is well-understood and  previously worked out explicitly
for many examples.  
See for instance \cite{KlassenMelzer},   and \cite{Corrigan} for  the affine Toda theories based on an arbitrary Lie group.  

    Let us illustrate with the so-called sinh-Gordon model, which is the 
affine Toda theory based on the Lie group  SU(2),   with potential 
\beq
\label{VshG} 
V(\phi) = 2 \mu \cosh (\sqrt{8 \pi}\, b \,\phi ),
\eeq
 where $\mu$  sets the energy scale  for the mass $m$ of the single particle in the spectrum, and $b$ is a dimensionless coupling.\footnote{Not to be confused with $b$ in 
 \eqref{betagcal0}.}   
   Here $\gcal = - 4 \pi \sin (\pi \gamma), ~~ \gamma \equiv b^2/(1+b^2)$,  and one finds  
\beq
\label{rhovacshG}
\rhovac = - \frac{m^2}{8 \pi b^2} \( 1 + b^2 + \frac{\pi^2}{6} b^4 + \ldots \).
\eeq
The above result can also be derived in Feynman diagram perturbation theory \cite{DestrideVega}.  
The leading term is non-perturbative in the coupling $b$ and the terms in parentheses correspond to perturbative corrections.
Indeed one recognizes $\zeta (2) = \pi^2/6$ which is commonplace in  perturbation theory,  and one can easily check the higher orders in $b$ involve 
$\zeta(2n)$ for integer $n$.

\def\phibar{\bar{\phi}}

 As a warm-up exercise  for 4 spacetime dimensions,  it is instructive to understand the leading $1/b^2$ term in \eqref{rhovacshG} without relying on integrability.  
 To this end we expand the $\cosh$ potential in \eqref{rhovac2D} in powers of $\phi$:
 \beq
 \label{shGPhi4}
 V(\phi) = 2 \mu + \frac{m^2}{2} \phi^2 + \frac{\lambda}{4!} \phi^4 + \CO( \phi^6 ),  ~~~~{\rm where} ~~~ m^2 = 16 \pi \mu b^2, ~~~ \lambda = 128 \mu \pi^2 b^4 \,. 
 \eeq  
Minimizing the potential $\d_\phi V(\phi) =0$,   in addition to a minimum at $\phi=0$ there is a minimum for  non-zero
$\phi = \phibar$ where  $\phibar^2 = - 6 m^2/\lambda$.  Ignoring for the moment that this $\phibar$ is only real for negative $\lambda$, 
one finds 
\beq
\label{shGPhi4b}
\rhovac = V(\phibar) = 2 \mu - \frac{3}{2}  \frac{m^4}{\lambda} = -\mu = - \frac{m^2}{16 \pi b^2} \, ,
\eeq
where we have used the expressions for $m^2$ and $\lambda$ in \eqref{shGPhi4}.    This agrees with \eqref{rhovacshG} up to a factor of $2$,   the discrepancy 
arising from the fact that we truncated the $\cosh$ potential to order $\phi^4$.   One can obtain the exact leading term with a similar semi-classical calculation
starting from the $\cosh$ potential \eqref{VshG},   where the minimum is at $\phibar = 0$ such that $V(\phibar) = 2 \mu$.    One delicate point is that in the 
thermodynamic Bethe ansatz the scalar particle corresponding to the field $\phi$ must be treated as a fermion since in the high energy limit the 
S-matrix $S= -1$.     Thus  $\rhovac = - 2 \mu$ which equals the leading term in \eqref{rhovacshG}.      It is important to note that the value of $\rhovac$  in
 \eqref{shGPhi4b} does not require spontaneous symmetry breaking (SSB).

 Let us now turn to 4 spacetime dimensions,   where one does not have the powerful  tools  of integrability available.
  In recent work \cite{ALrhovac1,ALrhovac2}  we formulated a  proper  computation of $\rhovac$ which renders it finite and well-defined
in 4 dimensional Minkowski space and its theoretical basis is  the same as  in two spacetime dimensions.  
  In \cite{ALrhovac1} we showed how to calculate $\rhovac$ from  QFT at finite temperature $T$,   which parallels the 
its computation from the thermodynamic Bethe ansatz in $d=1$.   Riemann's zeta function $\zeta (s)$ and its fundamental properties such 
as the functional equation that relates $\zeta (s)$ to $\zeta (1-s)$ played an essential role in the analysis in \cite{ALrhovac1},  and is analogous to modularity in $d=1$.        
    Namely,   let $\mzeron$ be the  fundamental renormalized scale of physical particle masses where $\mzeron$ represents the lightest mass particle,  and 
 $c(r)$ where $r = \mzeron c^2/k_BT$ be the scaling function such that the free energy density is given by 
 \beq
 \label{rhovacThermo}
 \CF = - \frac{\pi^2T^4}{90}  \, c(r),    ~~~~~~ {\rm with} ~~ c(r) = c_{\rm uv} + c_4 \, r^4 + \ldots
 \eeq
 for some constant coefficients $c_{\rm uv}, c_4$.\footnote{In 2 spacetime dimensions $c_{\rm uv}$ is the effective
 Virasoro central charge. In 4 spacetime dimensions,   $c_{\rm uv}$ is known from conformal field theory  if the theory is asymptotically free,  as in QCD. }      Then 
 \beq
 \label{rhovacC4}
 \rhovac = - c_4 \pi^2 \mzeron^4/90.
 \eeq    
 This was used in \cite{ALrhovac1} to estimate $\rhovac$ for QCD based on finite temperature lattice 
 results where $c_{\rm uv}$ is known since QCD is asymptotically a free conformal field theory.     
 
 One can also in principle compute $\rhovac$ directly from the S-matrix at zero temperature using the form-factor bootstrap \cite{ALrhovac2}.
 One advantage of this formulation  that all masses are the physical (renormalized) ones once the 2-particle form factor is properly normalized.   
  The form-factor bootstrap relates the two particle form-factor to the zero particle one,  which is a 1-point vacuum expectation value of the field
in question.    As in two spacetime dimensions,  we take  $\mzeron$ to be the physical mass of the lightest particle,   since in principle the  S-matrix and form factors for other particles can be obtained by bootstrapping this  lightest mass particle.   This is where we need {\it Assumption 1} stated in the Introduction,   namely that all particle excitations are coupled.    In connection with this,  in \cite{ALrhovac2} we stated the principle of {\it particular democracy},   wherein {\rm any} particle can be used to 
probe the Vacuum and its energy density,  and they should give the same value for $\rhovac$,   and this is known  to be true for the $d=1$  models considered there. 
    Applying this to $\langle 0| T_{\mu\nu} |0\rangle$,   this leads to a determination of the 
coupling $\gcal$.       As usual,  defining $\CT$ from the S-matrix $S = 1 + i \CT$,   then 
\beq
\label{rhovacFF}
\lim_{s \to \infty} \CT (s) = \frac{\gcal}{\mzeron^2} \inv{s},
\eeq
where $s=(p_1 + p_2)^2$ is the Mandelstam variable for 2-particle scattering.   It was then shown in \cite{ALrhovac2} that this leads to 
\eqref{rhovacg} which we repeat here
\beq
\label{rhovacg2}
 \rhovac =\frac{3}{4}  \frac{c^5}{\hbar^3}  \frac{\mzeron^4}{\gfrak} \, ~~~~~~~~~~(d=3).
 \eeq
For general spatial dimension $d$,    the factor of $3/4$ equals $d/(d+1)$ which agrees with \eqref{rhovac2D}. 
The derivation of \eqref{rhovacg2}  presented  in \cite{ALrhovac2}  from the form-factor bootstrap was new,  even in two spacetime dimensions.   In this derivation of  the above formula for $\rhovac$,  
since it is based on the properly normalized 2-particle form factor,   $\mzeron$ is the {\it physical} renormalized mass.

 For integrable theories in $d=1$,     the two above definitions of $\rhovac$, equations \eqref{rhovacC4} and \eqref{rhovacg2},   were shown to agree exactly for a 
 wide variety of integrable quantum field theories  \cite{ALrhovac1,ALrhovac2}.     Various consistency checks on 
  this formula were made,  including  that it must vanish for supersymmetric theories and also theories with a fractional supersymmetries where the hamiltonian is in the center of the universal enveloping algebra of the conserved charges.  (See \cite{ALrhovac1} and references therein.)       The coupling
 $\gcal$ can be positive or negative,   and $\rhovac$ can even oscillate in sign for a given model as a function of $\gcal$,  for instance for the sine-Gordon model where it oscillates between  positive and  negative,   and  is zero at the (fractional) supersymmetric points.  For the sine-Gordon model,  there is a special value of the 
 coupling $\gcal$ where the fractional supersymmetry is just $\CN = 2$ supersymmetry and $\rhovac$ vanishes there as it should   \cite{ALrhovac1} .            
  In 4 spacetime dimensions 
 we did not  rigorously prove  that equations \eqref{rhovacC4} and  \eqref{rhovacg2}  are equivalent definitions of 
 $\rhovac$,  thus \eqref{rhovacg} should still be viewed as  well-motivated but still conjectural.     It's important to note that the above arguments for the formula 
 \eqref{rhovacg} do not rely on any kind of spontaneous symmetry breaking.      
  For the remainder of the article we simply assume   the formula \eqref{rhovacg}.      
 
 In 4 spacetime dimensions,   we expect that the ``coupling" $\gcal$ has a renormalization group flow,  based in part on the log's in \eqref{rhoWein2}, 
 and we will assume this in the next section.
 One  can motivate this here by considering for instance  $\lambda \phi^4$ theory where $\gcal = \lambda$ to lowest order in perturbation theory.       Based on \eqref{shGPhi4b},   one expects that to leading order 
 $\rhovac \propto  \mzeron^4 / \lambda$.   It is known that the marginal coupling $\lambda$ has a non-trivial RG flow \cite{Peskin}.   
 To 1-loop order,
 \beq
 \label{betalambda}
 \mu \d_\mu \lambda =  \frac{b}{2 \pi} \,  \lambda^2 , ~~ ~~~~b= \frac{9}{8\pi} ,
 \eeq
 where increasing the energy scale $\mu$ corresponds to a flow to higher energies.     In this model,   $\lambda$ grows in the flow to high energies,
 which is to say it is a marginally {\it irrelevant} coupling.

 Independent  support for the formula  \eqref{rhovacg} comes from Swampland ideas \cite{Festina1,Festina2},  however this is rather  indirect  since the 
 arguments are very different as  they rely on black holes.     
 By studying a charged particle of mass $m$ in the presence of a black hole,   it was argued that 
 \beq
 \label{FestinaEqn}
  \rhovac \leq  \frac{m^4}{2 e^2} 
  \eeq
  where here  $\gcal$ is identified with the  quantum electrodynamic (QED)  fine structure constant $\alpha = e^2/(4 \pi \hbar c)$.
  Actual QED will not play a role in this article since we assume the the particle with mass $\mzeron$  is electrically neutral,
     however, as we will see,   it will be instructive to compare couplings and renormalization group parameters with QED,   just  to check if they are reasonable.      
  Our result \eqref{rhovacg} is stronger than \eqref{FestinaEqn}  since it is not an inequality,  but it is still consistent with \eqref{FestinaEqn}.

\subsection{Effective RG flow for Newton's constant based on the RG flow for $\rhovac$.}

The formulation of a gravitational Casimir effect in  Section  II  is independent of the detailed properties of $\rhovac$.     In this section we use the expression 
\eqref{rhovacg} to further develop our \darkuniverse \,  model  consisting of only vacuum energy.          
 For simplicity of subsequent discourse let us rewrite equations \eqref{Fried1d},\eqref{Fried2d} explicitly for $d=3$: 
\beq
\label{Friedmann}
\frac{\addot}{a} =  \(\frac{\adot}{a}\)^2 =   \frac{8 \pi G_N}{3} \rhovac .
\eeq

In the expression \eqref{rhovacg} for $\rhovac$,   by dimensional analysis $\gcal$ is a dimensionless coupling constant.   Based on the above arguments,   $\mzeron$ is the physical renormalized mass  of the lightest particle and we thus ignore its potential RG flow. 
  Also based on arguments of the last sub-section,    we  expect that $\gcal$ has an RG flow,   based on the comments in  \cite{ALrhovac1,ALrhovac2,LeClairUniverse} and  equation \eqref{FestinaEqn} where $\alpha = e^2/4 \pi \hbar c$ is the fine structure constant with a known RG flow coming from QED.   
If we associate $\gcal$ with a marginal coupling,   $\gcal$ could be marginally relevant or irrelevant.     
Let us assume the following beta function,   where $\mu$ is an energy scale,    and increasing $\mu$ corresponds to an RG flow to higher energies\footnote{The factor of $1/2\pi$ is motivated by standard definitions of fine-structure constants $\alpha$,   such as 
$e^2/(4 \pi \hbar c)$ for QED.      
If $\gcal =   \alpha = \frac{g^2}{4 \pi \hbar c }$ where $g$ is a gauge coupling with beta-function 
 $\beta_g  =  \frac{dg}{d \log \mu}  =  b \frac{g^3}{16 \pi^2}$,   then
$\beta_\gcal  =  b \gcal^2/2\pi$.  For pure QED with $N_f$ species of Dirac fermions,  $\gcal_0 \approx 1/137$ and  $b=4N_f /3$.    For $\lambda \phi^4$ theory,   if $\gcal = \lambda$ then
$b= 9/8\pi$. See  \cite{Peskin}.}
\beq
\label{betagcal}
\mu \d_\mu \gcal =  \frac{b}{2 \pi} \gcal^2 .
\eeq
The above beta function is typically just the 1-loop approximation,    however we will argue below  based on cosmological data,  that the coupling $\gcal$ is presently small  so that a 1-loop approximation is justified.   Furthermore,   in well-understood QFT's such as QED,   the Landau pole discussed below  is not removed by higher order corrections to the beta function.    
We assume $\gcal >0$ for a positive $\rhovac$ so that we are dealing with de Sitter space.         
For $\gcal > 0$,   if $b$ is positive the marginal coupling $\gcal$ increases at higher energies,  namely it is marginally irrelevent,     whereas if $b$ is negative $\gcal$ flows to zero in the UV,  i.e. is asymptotically free.    
Let $\gcal_0 = \gcal (\mu_0)$ where $\mu_0$ is the cosmological energy scale today at time $t_0$.    
Integrating the beta function is straightforward:   
\beq
\label{gofmu}
\frac{\gcal (\mu)}{\gcal_0}  = \inv{1 - \bhat    \log (\mu/\mu_0) } ,  ~~~~~ \bhat \equiv  \frac{b \,\gcal_0}{2 \pi} .
\eeq
The RG parameter $\bhat$ is the primary new constant for applications to cosmology in our framework.      
As we will see,   based on cosmological data,   we will  propose $\bhat >0$,  corresponding to 
a marginally irrelevant coupling,  and unless otherwise stated,  below $\bhat$ is positive. 
     
The pole at $1-\bhat \log(\mu/\mu_0) = 0$ is commonly referred to as a Landau pole,  and signifies 
the theory is not UV complete.    
 It will play an important role below where  we will return to addressing  its  significance.  
 Landau poles are commonplace in elementary particle physics and condensed matter physics,   since they are generic to marginally irrelevant perturbations.   
See for instance \cite{LandauPole1,LandauPole2,LandauPole3} and references therein.    
Although QED itself is not relevant to this article,           
  let us mention that it also  is not UV complete due to a Landau pole,   however in this context it occurs at much  higher energies than our proposed $\zmax$,  and is thus commonly viewed as a purely  academic issue,  and is furthermore complicated by chiral symmetry breaking.      (See footnote 17).

From the expression \eqref{rhovacg} one has 
\beq
\frac{\rhovac (\mu)}{\rhovac(\mu_0)} = \frac{\gcal_0}{\gcal(\mu)} ,
\eeq
since we assume $\mzeron$ is already renormalized to the physical mass.   
Thus $\rhovac$ in \eqref{Friedmann} should be replaced by $\rhovac (\mu)$.       In order to express this equation in terms of 
$\rhovac = \rhovac (\mu_0)$ today,   it is  convenient,  and meaningful,  to incorporate the RG flow into an induced flow for Newton's constant:   
\beq
\label{FriedmannCG}
  \( \frac{\adot}{a} \)^2 = \CH(t;\mu)^2 \equiv  \frac{8 \pi \CG (\mu) }{3} \rhovac (\mu_0),   
\eeq
where 
\beq
\label{CGdef}
\CG (\mu) = \CG (\mu_0 )  \frac{\gcal_0}{\gcal (\mu)} =  \CG (\mu_0 ) \( 1 - \bhat  \log (\mu/\mu_0) \).  
\eeq
It is meaningful to incorporate the RG flow for $\rhovac$ into an induced flow for $G_N$ since 
we view $G_N$ as being determined in the far future where vacuum energy dominates and only afterwards incorporate adding matter and radiation  as we will do below.   
If one interprets $\CG (\mu)$  in \eqref{CGdef}  as an energy scale dependent Newton's constant,   then one sees 
that for $\bhat >0$,  {\it the strength of gravity decreases as one increases the energy scale $\mu$}.  At extremely low energy scales $\mu$,   $\CG(\mu) \to \infty$.   
The quantity $\CH(t;\mu)$ is interpreted as the solution to Einstein's equations with a Newton's constant
$\CG (\mu)$ that depends on the energy scale $\mu$.    
Then based on \eqref{Friedmann} one has 
\beq
\label{HubT1}
\frac{\CH(t;\mu )}{\CH(t; \mu_0)} = \sqrt{\frac{\CG(\mu)}{\CG(\mu_0)}} = \sqrt{1 - \bhat  \log (\mu/\mu_0) } .
\eeq

\section{Exact solution to the scale factor $a(t)$ for the solely  \darkuniverse}

The natural energy scale $\mu$ in cosmology is based on its  temperature $T$.   In the $\Lambda{\rm CDM}$ model this temperature is
tracked by the cosmic microwave background (CMB).      Thus far we have only considered vacuum energy with no radiation nor matter,  and will 
turn  to the  $\Lambda{\rm CDM}$ model below.    As a warm up exercise,    let us assume the standard relation between temperature and the
 scale factor:\footnote{For the $\Lambda{\rm CDM}$ model,  
 $T_0 = 2.7 K$,  thus $\mu_0 = k_B T_0 = 2.35 \times 10^{-4} \, {\rm eV}$.}
\beq
\label{Tofa} 
T(t)  = \frac{T_0}{a(t)}  ,
\eeq
since as we will see below the main features will persist with the addition of matter and radiation.   
For our model of pure vacuum energy,  the temperature $T$ can be thought of a generic heat bath with the Planck black body spectrum.   
In fact in Section III we argued that de Sitter space has a temperature and we estimated this very low temperature $\Tinf$ in the far future
where the evolution is dominated by vacuum energy density \eqref{HawkingT}.
As is also standard,  we express $a(t)$ in terms of the redshift $z$,  $a(t) \equiv  1/{(1+z(t))} $,   such that 
\beq
\label{muz}
\mu/\mu_0  =  T/T_0 =  1+z(t).  
\eeq
This scaling is expected to hold  at least post Big Bang Nucleosynthesis,   which is roughly a few minutes after the hot big bang.
Since our model thus far only consists of vacuum energy density $\rhovac$,    we will  assume the equation \eqref{muz} is valid for all times.    
We thus identify $\mu_0 = k_B T_0$,  
and  $z=0$ with $\mu_0$.
If we identify $G_N = \CG (\mu_0)$,  then 
\beq
\label{CGdef2}
\CG (\mu) = \CG (z) =  G_N  \( 1 - \bhat  \log (1+z) \) = G_N  \( 1 + \bhat  \log (a) \) .
\eeq
Note that $G_N =0$  when $a = \amin = e^{-1/\bhat}$.    For smaller $a(t)< \amin$,  
in principle $\CG(\mu)$ could change sign,   which would signify a transition to anti-de Sitter space,   however as we will see,  $a(t)$ never drops below $\amin$ and the logarithmic branch cut 
from $a=0$ is never reached.

Turning to \eqref{FriedmannCG},    $\adot/a=H$ is no longer constant since $\mu$ is time dependent according to \eqref{muz}.   Taking the square-root we wish to solve 
\beq
\label{SqRootFriedmann}
\( \frac{\adot}{a} \) = \pm   \sqrt{1 + \bhat \log a(t) } ~ \( \frac{8 \pi G_N  \rhovac(\mu_0)}{3} \)^{1/2}  .
\eeq
 This  additional kind of time dependence does not originate directly  from Einstein's equations but rather from the energy scale $\mu$ dependence of Newton's constant and 
$\mu$  being tied to the temperature through equation \eqref{muz}.    By analogy,     the classical Maxwell's equations by themselves cannot account for the well-known energy scale dependence of  fine structure constant $\alpha = e^2/(4 \pi \hbar c)$,    which implies it can depend on temperature,   and if the temperature is time dependent,   such a time dependence is not captured by the time dependence of Maxwell's equations.     
This leads to a significant and interesting  modification of the de Sitter space solution to $a(t)$.   
Let us first chose the positive sign $+$ on the RHS of  
\eqref{SqRootFriedmann} such that $\adot >0$ in the future.   As we will show,  choosing the minus sign  will be relevant for the far past,  where the solutions match at
$a(t) = \amin$.                 The equation \eqref{FriedmannCG} can be 
explicitly integrated,  yielding a  ``inverted gaussian"  correction to  de Sitter space,  which we will simply refer to as gaussian de Sitter space:
\beq
\label{ModdeSit1}
a(t) = e^{-1/\bhat} \,  \exp \[  \inv{\bhat}  \(  \frac{\bhat H_0}{2} (t-t_0)+1 \)^2    \] 
=  \exp \[ H_0 (t-t_0) +  \bhat H_0^2  (t - t_0 )^2 /4\] .
\eeq
Above,  $H_0$ equals $\adot/a$ at the present time $t_0$,  since the constant of integration is chosen 
such that $a(t_0) =1$.    One sees that near the present time,  namely $t-t_0$ small,  the second term in the exponential $\propto (t-t_0)^2$ is suppressed,   and if $\bhat =0$ one recovers the usual de Sitter solution $a(t) = e^{H(t-t_0)}$ with $H = H_0$.     

The most interesting feature of the gaussian de Sitter solution \eqref{ModdeSit1} is that there is no singularity $a=0$ for all times.   In spite of the  Landau pole,  the  solution is still regular,  since the argument of the square root,  $1 + \bhat \log a(t)$,  is always positive since $a(t) > \amin$ for all times.  
 This  minimum value of $a(t)$ is 
\beq
\label{amin} 
a(t) > a_{\rm min} \equiv e^{-1/\bhat} ~~\forall t  ~~~~~\Longrightarrow ~~ 
 \zmax = \inv\amin -1 = e^{1/\bhat} -1 .
\eeq
The time $\tmin$ where $a(\tmin) = a_{\rm min} $ is 
equal to 
\beq
\label{tmin} 
\tmin = t_0  - \frac{2}{H_0 \bhat} .
\eeq
Note that as $\bhat \to 0$,   $\tmin \to -\infty$ as expected for de Sitter space.  
In the next section we will approximate $\bhat \approx 0.02$ based on the so-called
Hubble tension,  which leads to a very large, but {\it finite}  value for $\zmax$ (see eq. \eqref{zmaxValue} below.)

Let us now turn to the far past $t \to -\infty$.   In ordinary de Sitter space,
$\lim_{t\to -\infty}  a(t) =0$,   although it is known this is not a true geometric curvature  singularity.  
For our gaussian de Sitter space,    the solution  \eqref{ModdeSit1} is formally  still valid for $-\infty < t <\tmin$,  and thus valid for all times.     
In fact
 \beq
\label{minustsymmetry0}
\lim_{t \to \pm \infty}  a(t) = \infty .
\eeq
This is a consequence of a stronger relation between the past and future.     One can see from the solution 
\eqref{ModdeSit1} that it has the symmetry
\beq
\label{minustsymmetry}
a(t) = a(-t + 2 \tmin) .
\eeq
It is important to note that the above symmetry is independent of the present time $t_0$.     
At the time $\tmin$,  ~   $a(\tmin)=\amin$ is self-dual.     Recall that at the time $\tmin$ with scale factor $\amin$,  
the effective Newton constant $\CG (\mu)$ vanishes such that the solutions of \eqref{SqRootFriedmann} with $+$ verses $-$ agree,    thus for $t< \tmin$ one should take the solution with the minus sign,   
where $a(t)$ {\it decreases}. 

To summarize,    the solution \eqref{ModdeSit1} is valid for all times $-\infty < t < \infty$ where 
$a(t)> \amin$.      The complete history of such a model universe,  which we referred  to as the \darkuniverse,   is that in the far past $t\to -\infty$  the scale factor $a(t)$ is infinitely large and starts to compress  down to $\amin$ which occurs  at a time $\tmin$.    Beyond this time,   $t> \tmin$,    the universe expands until the scale factor $a(t)$  is again  infinite at
 $t= +\infty$.    This takes an infinite amount of time,   and  avoids any geometric singularities at $a=0$.    There is no ``Big Bang"  corresponding to scale factor $a(t)=0$.  
 Rather,  at time $\tmin$  the universe smoothly transitions from a compression to an expansion,   and $\tmin$ is just the  time when the universe is hottest.    The Big Bang is now better  described as a  Big Swing. Henceforth,   by ``big bang" we refer to the time $\tmin$ which is a very  hot big bang and not associated with any curvature singularities at $a=0$.     
     A picturesque analogy is  the harmonic  pendulum with the bottom of the swing corresponding to $\amin$,   and at the present time the universe is on the up swing.  
 When we add quantum matter and radiation in the next section,   these are viewed as excitations above the vacuum,  and as the pendulum is on the downswing it accumulates kinetic energy which can  excite 
 these states. 
         Based on \eqref{minustsymmetry},  a consistent possibility is that this process then repeats itself,  such that  this universe is in a sense oscillating about its vacuum like a pendulum harmonic oscillator about its ground state.      At the top of the swing the universe 
 is at its largest.             If this is the case,  then the age of this gaussian de Sitter  universe is  even more  eternal.

\section{Adding matter:  the $\LambdaCDM$ model re-examined and the Hubble Tension}

Thus far we have only considered vacuum energy density $\rhovac$ which led to the exact solution of the \darkuniverse ~presented above.       We gave a heuristic derivation of Einstein's equations in this  context from the gravitational Casimir effect we formulated above.       The Universe also contains radiation and matter,   which we view as excitations above the vacuum,   which must be included in $T_{\mu\nu}$.
The standard Friedmann equations have three sources,   as in \eqref{LCDMHH} below,  but without the  $(1+ \bhat \log a)$ factor.    The question naturally arises:
should the renormalization properties of $\rhovac$ be kept to the $\OmegaLambda$ only, i.e. the $\rhovac$  term,   or should it be incorporated also in the 
$\Omega_{\rm m}$ and $\Omega_{\rm rad}$ terms?      We take the following point of view,   although it perhaps requires more scrutiny,   or at least a more complete
argument.    
The \darkuniverse~   is considered a skeleton of our Universe with pure vacuum energy $\rhovac$  and this led to the emergence  of $G_N$ from $\rhovac$.  
Matter and radiation are excitations over the vacuum,   however we continue to assume that $G_N$ is determined by vacuum energy since we argued that the latter is
the origin of Gravity itself.            
In our analysis of the  \darkuniverse,   the renormalization group flow coming from $\rhovac$ was absorbed into an energy scale  $\mu$ dependent  Newton's constant $G_N$.    If the energy scale $\mu$ is  fixed,     then we add matter and radiation
by demanding  local energy-momentum conservation,  which  requires vanishing of the  covariant divergence  $\nabla^\mu T_{\mu\nu} = 0$,  which is automatic due to the Bianchi identities.   Thus,   for instance,  Newton's universal law for the gravitational force between two masses at low energies is subsumed as a consequence.          At this point one must deal with  
 the usual Einstein's equations except with an energy dependent  Newton's constant $\CG(\mu)$.      Based on this we propose  that:    
\beq
\label{LCDM}
\( \frac{\adot}{a} \)^2  = \frac{8 \pi}{3} \CG(\mu) \, \rho_{\rm total},
\eeq
with $\CG (\mu)$ defined in \eqref{CGdef2}.      
It is conventional  to express \eqref{LCDM}  as follows 
\beq
\label{LCDMHH} 
H^2  \equiv \( \frac{\adot}{a} \)^2  = H_0^2  \(1+ \bhat \log a \) \(  \frac{\Omega_{\rm rad}}{a^4}  +   \frac{\Omega_{\rm m}}{a^3}  + \OmegaLambda \), 
\eeq
where by definition $a(t_0) = 1$ at the present time $t_0$,  $H(t_0) = H_0$,  and $\Omega_\Lambda$ is the $\rhovac$ contribution.   
As for the \darkuniverse,     the additional time dependence due to the factor  $(1+ \bhat \log a(t))$ does not follow directly  from the Einstein equations,   
but rather from the energy scale $\mu$ dependence of Newton's constant and 
$\mu$  being tied to the temperature through equation \eqref{muz}.    The Bianchi identities are only valid locally where $\mu$ is a fixed and thus Newton's constant 
$\CG (\mu)$ is time independent.     
  The zero curvature ($k=0)$ motivated in Section II implies 
$\Omega_{\rm rad} + \Omega_{\rm m} + \OmegaLambda = 1$.    The above prescription \eqref{LCDMHH} for dealing with the issue raised preserves  the 
minimal scale factor $\amin$ as we will show below.    It is also a key aspect of our proposal for dealing with the Hubble tension. 
   
In the $\LambdaCDM$ model one ignores radiation,  i.e. $\Omega_{\rm rad} \approx 0$,   as this is a well-justified approximation during this epoch.     
There is a deepening discrepancy in the $\LambdaCDM$ model based on relatively recent  astrophysical measurements referred to as the ``Hubble tension".      
The ideas in this article offer a potential resolution which we now present.       As usual,   let $H_0=\adot/a$ be the Hubble constant {\it today}. 
It should be a fixed constant regardless of the manner in which it is measured.    The discrepancy arises from two very  different kinds of 
measurements.     The first comes from ``local" measurements based on Type Ia supernovae,   where typical redshifts are relatively low, 
in the range $z=0.02–0.15$ (SHOES \cite{SHOES}.).       The other determination of $H_0$ is based on  CMB 
data   at much higher redshift $z=1100$.    (Planck \cite{Planck}).    The two contradictory values currently reported are 
\beq
\label{Hos}
 \HoCMB = 67.4 \pm 0.5  ~ {\rm km/s/Mpc}, ~~~~~
 \HoSupernova = 73.0 \pm 1.0 ~ {\rm km/s/Mpc} ,
 \eeq
 which differ significantly enough for some prominent  researchers   to question the $\LambdaCDM$ model,   with suggestions that this could signify beyond standard model physics \cite{HubbleTensionRef,Riess2}.   In the latter it was suggested that this could be due to a variable Dark Energy component,  but no specific theoretical model was 
 advocated.   

Our interpretation of the discrepancy in these two values for $H_0$ in \eqref{Hos} is the following. 
Consider an observer making measurements of $H(t)$ at an earlier time,   where their  measurements are in a small range of  higher redshift  $z$ than today where today $z=0$.    At this epoch,   $\CG (\mu)$ is effectively lower  according to \eqref{CGdef2}.     If this observer fits their  data to \eqref{LCDMHH} they  will predict a lower value of $H_0$ compared to a fit based on data 
taken at $z \approx 0$.      The CMB measurements probe the state of the Universe at this earlier time,    thus though CMB measurements are made at the present time $t_0$,   they reflect a  hypothetical observer making measurements at an earlier epoch where $z\approx 1100$.   
We thus propose  based on \eqref{HubT1}
\beq
\label{HubT2}
 \frac{\HoCMB}{\HoSupernova} \approx  \sqrt{1- \bhat \log (1+z)},  ~~~~ {\rm for} ~z=z_{\rm CMB} =1100 ~~\Longrightarrow ~~ \bhat \approx 0.02
\eeq
where we have set $z\approx 0$ for the supernovae measurements.  
Based on the measured values in \eqref{Hos},    the above leads to $\bhat \approx 0.02$.

Let now turn to explicitly solving \eqref{LCDMHH} in ``real time",  namely with $\CG (\mu(t))$ given in 
\eqref{muz},  as we did for pure vacuum energy with the result \eqref{ModdeSit1}.   
Before even integrating the equation \eqref{LCDM},   one can immediately see that  there are no real solutions unless 
$1 + \bhat \log a > 0$.    Thus there is still a minimal  scale factor $a(t)>\amin$ which is the same as in \eqref{amin} and attributed to the Landau pole.     This is a robust conclusion,   regardless of whether $\rho_{\rm rad}$ is incorporated as the dominant form of energy in the very early Universe.     The above value for $\bhat$,   leads to a very high 
value for the maximal red shift and temperature:\footnote{This is not an unreasonable  value for $\bhat$,  in that 
$\gcal_0 = 2 \pi \bhat/b = 0.12/b$,   and typically  $b = \CO (1)$.    For comparison,   for QED   $b=4N_f/3$ for 
$N_f$ species of Dirac fermions,  and if $\gcal_0 = 1/137$ then    $\bhat = .002 N_f$ which is about 10 times smaller than in \eqref{zmaxValue}  and $\zmax\approx 10^{500}$ is far beyond the Planck scale.}
\beq
\label{zmaxValue}
\bhat \approx  0.02 ~~~ \Longrightarrow ~~ \amin \approx 2 \times 10^{-22} , ~~~~ \zmax \approx 5 \times 10^{21} ~~~\Longrightarrow ~~ ~\Tmax \approx 10^{22} \, K.
\eeq
This is far above what has been directly probed by the CMB,  the latter being  $z_{\rm CMB} = 1100$.   
Thus our proposed RG flow for Newton's constant implies only small logarithmic corrections to the 
$\LambdaCDM$ model.      For instance it doesn't  significantly alter  what was previously considered as the age of the Universe,  as we will show  below.

To further justify the last statement,   let us turn to the $\LambdaCDM$ model where radiation $\Omega_{\rm rad} = 0$ is a good approximation.   
The scale factor $a(t)$ can be solved numerically starting from \eqref{LCDMHH}.      
In order to probe potential singularities at early times where the matter dominates due to the $1/a^3$ in \eqref{LCDM},   let us neglect the $\OmegaLambda$ term as an approximation,   since the resulting equation can be solved analytically and leads to more transparent conclusions.            
Then one wishes to integrate the equation 
\beq
\label{LCDM2}
\( \frac{\adot}{a} \)^2 = H_0^2  \,\(1 + \bhat \log a \)\,  \frac{\Omega_m}{a^3} .
\eeq
When $\bhat = 0$,  this is easily integrated:  
\beq
\label{Nobhat1}
a(t)^{3/2} = \tfrac{3}{2}  H_0 \sqrt{\Omega_m}  \( t - t_0 + \frac{2}{3 H_0 \sqrt{\Omega_m}} \),
\eeq
where $a(t_0) = 1$ is an initial condition.   
There exists a time $\tmin$ where $a$=0:\footnote{In the cosmology literature the  convention is to 
shift $t \to t + \tmin$ such that $a =0$ at the shifted time $t=0$.  The difference  $t_0 - \tmin$ essentially  represents the age of the Universe if one assumes the Universe came into existence at time $\tmin$,  which is questioned in this article.}
\beq
\label{Nobhat2}
a(\tmin) = 0 ~~{\rm for} ~~ 
 \tmin = t_0 - \frac{2}{3 H_0 \sqrt{\Omega_m}} .
\eeq
For $\bhat \neq 0$ the equation  \eqref{LCDM2} can also be explicitly integrated.    As for the pure vacuum energy case of the last section,   taking the square-root of    \eqref{LCDM2} introduces a $\pm$ sign as 
in \eqref{SqRootFriedmann},   where $+/-$ corresponds to the far future/past.     For the $+$ sign 
the solution $a(t)$  can be  expressed  implicitly in terms of the imaginary 
error function  ${\rm erfi }(x) \equiv   {\rm erf} (i x)/i$: 
\beq
\label{ErfiEq}
\erfi \sqrt{ \frac{3}{2\bhat} \( 1 +  \bhat \log a(t) \)} =   (t-t_0) H_0 \,\, 
 \sqrt{ \frac{ 3 \bhat \, \Omega_m }{2 \pi} } \, e^{3/2\bhat} + \erfi \sqrt{\frac{3}{2 \bhat}}  ,
 \eeq
 where again we have imposed $a(t_0) =1$.    
The argument of the $\erfi$ function on the LHS must be real otherwise the $\erfi$ function is imaginary.      
Thus due to  the square-root branch cut 
$\sqrt{1 + \bhat \log a}$  inside $\erfi$,   there is a minimal value of $a(t)$ again given by \eqref{amin}.   
It is not difficult to check that the same conclusion is reached in a radiation dominated era where the $1/a^3$ term is replaced by $\Omega_{\rm rad} /a^4$.    
This minimal value of $a(t)$ occurs at a time $\tmin$ where  $a(\tmin) = \amin$,   which is determined by 
when the LHS of the above equation is zero since $\erfi (0) = 0$: 
\beq
\label{tminLCDM}
a(\tmin) = \amin = e^{-1/\bhat} 
\eeq
where $\tmin$ is determined by  the equation 
\beq
\label{tminLCDM2}
t_0 - \tmin   =\inv{H_0}  e^{-3/2\bhat} \sqrt{ \frac{2 \pi}{3 \bhat\, \Omega_m}} \erfi \sqrt{\frac{3}{2 \bhat}} 
=  \frac{2}{3 H_0 \sqrt{\Omega_m}} \( 1 + \tfrac{1}{3} \,\bhat +  \tfrac{1}{3} \,\bhat^2+ \tfrac{5}{9} \, \bhat^3  + \ldots  \).
\eeq
The solution \eqref{ErfiEq} is again  valid for all times $-\infty < t < \infty$ due to a symmetry  between the far past and  far future.   Using $\erfi(-x) = - \erfi(x)$,   one can show  that   equation \eqref{minustsymmetry},  
which equates $a(t)$ with $a(-t + 2 \tmin)$,
 remains valid with $\tmin$ given in 
\eqref{tminLCDM2},   as does \eqref{minustsymmetry0}.  

Based on the above analysis,  
 let us comment on the possible implications for the very early universe in our model.     Henceforth,  by   ``early universe"  we mean 
 around the time $\tmin$,  where the universe is hottest and the expansion phase begins.   Recall in our model,   time $t$ actually extends to 
 $t= -\infty$,  and this is the earliest universe.        
 The salient  features of the evolution of $a(t)$ from $t= -\infty$ to $t=+\infty$ is essentially the same as 
 for the pure vacuum energy case,  i.e. the \darkuniverse~  described in the last paragraph of the last section,  and this justifies viewing the \darkuniverse ~ as
 a skeleton of our universe based only on the Vacuum.  
 Namely at $t=-\infty$,    $a(t) = \infty$ and starts to decrease until it reaches $\amin$ at time $\tmin$,   then begins to increase back to $\infty$ as $t \to +\infty$.      At no time in this evolution is $a=0$,  again avoiding geometric curvature singularities.        There is no {\it singular} Big Bang at any time,   rather 
 at time $\tmin$ the universe is hottest since the redshift $z$ is at its maximum.       The quantity 
 $t_0 - \tmin \approx 13.7$ billion years,  does not represent the age of the universe,  but rather the time elapsed since the scale factor was at its minimum.   The age of this model universe is actually infinite,   which we find appealing,   since the universe does  not originate from  an incomprehensible singularity at $a=0$.    As described above  for the 
 \darkuniverse,   the time evolution of the universe is better described as a back and forth swing of a pendulum, 
 where the bottom of the swing occurs at $a(\tmin) = \amin$,    and we are currently on an upswing.     
  All the energy of the  hot universe at time $\tmin$ was accumulated during the downswing,   thus such a universe was {\it not created out of nothing},  since the
  Vacuum is not nothing;   it is something with a  ground state energy density that is a  source of energy,  if not the original source of all energy in the Universe.

The details of this time evolution of $a(t)$ are  very sensitive to the value of $\bhat$ due to the exponentials in 
\eqref{zmaxValue}.  For purposes of illustration,    let us assume $\bhat =0.02$ as inferred above  from the Hubble tension.    
First,  one sees from \eqref{tminLCDM2}  that for small $\bhat$ this amounts to a small corrections 
to $t_0 - \tmin$ which is normally associated to the age of the Universe.   
 If our estimate of $\bhat$ in \eqref{zmaxValue} is approximately correct,  
then this value for $\zmax \approx 5 \times 10^{21}$ is deeply into the radiation dominated era,   since the value of $z=z_{\rm eq}$ where matter and radiation are equivalent in density is only about $z_{\rm eq} \approx 3400$.    
At times much earlier than during the $\LambdaCDM$ era,    equation  \eqref{muz} 
may not be exact at all times,  since new degrees of freedom can be excited from the Vacuum and go through various phase transitions that modify  \eqref{muz} in relatively short
in time epochs.   However if we simply assume it  extends to the Planck scale with $T_0 =2.7$K,   then 
$z_{\rm planck}  \approx 5 \times 10^{31}$,   which is much higher than $\zmax$ by 10 orders of magnitude.     This  shows   that  the Planck scale $\ell_p$ plays no role in determining $\zmax$.     However a smaller value for $\bhat$ could  easily raise $\zmax$ up to the Planck scale,    where quantum gravity effects could modify our findings.

\def\zinflation{z_{\rm inflation}}

From  the  perspective on Gravity presented thus far,   
earlier times  $t<\tmin$ where  $a(t) < \amin$  do not exist in our model universe.   
    In the standard cosmology,  this 
earlier era is thought to be described by inflation.   Inflation is postulated 
as a way to get from the singularity at $a=0$ to a hot big bang, however it is not yet 
universally  accepted as being a complete  theory.   It requires invoking new fields such as the inflaton and fine-tuning their parameters for a ``slow-roll",  incorporating 
re-heating,  etc.     
   Very large numbers are naturally  involved  since the singularity at $a=0$ formally corresponds to $z=\infty$.       Typical  models require 
inflation to last for 50-60 e-foldings to inflate the Universe to the proper initial conditions from $a= 0$,
then transition to a re-heating scale to bring up the temperature to that of a hot  big bang,  
and this rapid expansion makes the Universe flat and homogeneous.    
The range of redshift $z$ over which inflation hypothetically occurs is strongly model dependent,  especially on what the re-heating temperature is.     The duration of inflation, marking its beginning and end,   can  range from 
$10^{22} < \zinflation  < 10^{53}$ depending on the model,   where the upper limit can even be significantly larger than
$z_{\rm planck}$.   It is noteworthy that at the lower end of this range based on lower
reheating temperatures,   namely where inflation ends at $\zinflation = 10^{22}$,  is close to our $\zmax = 5 \times 10^{21}$.    Since this inflationary epoch does not actually exist in our model due to the minimal scale factor $\amin$,   one should question whether inflation is really  necessary.         At $t = \tmin$  our model universe is already expanded since $a(t)> \amin$,   it is very  hot,  and is already flat with $k=0$.     In Section  II we  explained why the spatial curvature $k$ is naturally zero since we view Gravity as arising from vacuum energy in flat Minkowski space.   
The value of $\amin = e^{-1/\bhat}$   perhaps plays a role equivalent to hypothetical inflation models since 
it represents  roughly 50  e-foldings  since $e^{1/\bhat} \approx e^{50}$ for our estimate of $\bhat \approx 0.02$.  
On the other hand,  one must be careful not to discard the successes of inflationary models,   in particular their  hallmark prediction of quantum density fluctuations that 
seed future galaxies and has been observed in the CMB.    In our model universe,   such quantum fluctuations could arise from the field corresponding to the 
particle with mass $\mzeron$ itself,  which should be interpreted as doing the job of the inflaton.        Clearly this requires further study beyond the original scope of this article.

\section{Consistency with  various types of constraints and experiments}

In this section we compare the primary aspects of our model to known observations and experiments of various types.  
We will base our comparison based on our estimate  $\bhat \approx 0.02$,   however conclusions are rather sensitive to the value of $\bhat$ and could easily be 
accommodated with a smaller $\bhat$.

\subsection{$H_0 (z)$ trend for low $z$ Supernovae}

Since \eqref{HubT2} is only based on the two data points $z=0$ and $1100$,   more compelling support for our proposal would 
be a fit for a range of redshift $z$  that would confirm the trend proposed above: 
\beq
\label{HubT3}
 \frac{H_0 (z)}{H_0 (z=0)} =  \sqrt{1- \bhat \log (1+z)} = 1 - \frac{\bhat}{2} \, z + \frac{\bhat (2-\bhat)}{8} \, z^2 + \CO (z^3) .
\eeq
  This is possible with supernovae since they exist in a range of redshifts.
Remarkably this trend has been very recently observed for a large data sample of supernovae in the range of still relatively low $z$,    $0.001 <z <2.3$,  based on data 
analyzed  at the National Astronomical Observatory of Japan  \cite{Dainotti2021,Dainotti2022,Dainotti2024,Dainotti2025}. 
 In the latter work,   a fit to the functional  form 
 \beq
\label{HubT4}
 \frac{H_0 (z)}{H_0 (z=0)} \approx  \inv{(1+z)^\alpha}  = 1 - \alpha \, z + \frac{\alpha (1+ \alpha)}{2} \, z^2 + \CO(z^3) 
\eeq
was considered,   motivated by some versions of $f(R)$ modified gravity and other theoretical models which typically invoke additional fields,  for instance
Brans-Dicke scalars.  
(See  \cite{Montani,Montani2,Dainotti2021,Dainotti2022,Dainotti2024,Dainotti2025} and references therein.)   Although the two formulas \eqref{HubT3} and \eqref{HubT4} look rather different,
to lowest order in small $z$,   they agree precisely with $\alpha = \bhat/2$.      In  \cite{Dainotti2021,Dainotti2022,Dainotti2024,Dainotti2025} 
values of $\alpha \approx 0.01$ were reported,  which
 agrees with our estimate of $\bhat\approx 0.02$ inferred from the single  CMB data point $z=1100$ in \eqref{HubT2}.      This strongly suggests that the trend in $H_0 (z)$ seen in supernovae extends all the way to $z=1100$,  and this provides the strongest support for our model thus far.\footnote{We are currently working with Maria Dainotti's group at NAOJ to perform a detailed statistical analysis 
in order to try and distinguish between the two formulas \eqref{HubT3} and \eqref{HubT4}.   Preliminary results appear to validate our estimate of $\bhat \approx 0.02$ 
\cite{ALMaria}.}
   It clearly would be very interesting to extend this range of analysis to supernovae with higher $z$ or to gamma ray bursts.

\subsection{Known bounds on the time-variation of Newton's constant}

  If the energy scale varying $G_N$ proposed in this article is correct,    it is testable by other kinds of 
 measurements.
For cosmology,  our  formula \eqref{CGdef} for the scale dependent Newton's constant $\CG(\mu)$ implies that the effective Newton's constant is
 time dependent if one ties $\mu$ to the time-dependent scale factor as in
 \eqref{muz}.      Unfortunately,  presently there only exists   experimental bounds on the time variation of $G_N$,  usually reported
 as $\dot{G}_N/G_N$ today,   rather than non-null results.     Nevertheless,  let us check consistency with some known bounds.         
  Based on \eqref{CGdef2} one has the simple formula 
 \beq
 \label{Gdot}
 \frac{\dot{G}_N}{G_N} (t)   =  H(t)\cdot \bhat,
 \eeq
 where as above $H= \adot/a$.      
 At the present time $t_0$,  
 \beq
 \label{Gdot2}
 \frac{\dot{G}_N}{G_N}   =  H_0 \cdot  \bhat = 6.7 \times 10^{-11} \cdot  \bhat ~  /{\rm yr} ~~~~({\rm today}),
 \eeq
 where we have used $\HoCMB$.     Henceforth,  $\dot{G}_N/G_N$ refers to the present time.  
 
 \bigskip
 \noindent
 {\it Pulsar timing and gravitational waves.} ~
The strongest experimental constraints come from pulsar timing \cite{Taylor} based on pulsar observations spanning 30 years \cite{Taylor}.    Over this rather narrow window,  
one can constrain  $ \dot{G}_N/G_N  < 2 \times 10^{-12} /{\rm yr}$.   
   These pulsars exist at relatively low redshift $z$.   Thus based on  equation \eqref{Gdot2} with $\bhat =0.02$, 
   $\dot{G}_N/G_N  \approx  1.4 \times 10^{-12} \, /{\rm yr}$ which is very close to the bound based on pulsars.     
LIGO provides weaker  constraints  $ < 10^{-9}/{\rm yr}$  \cite{LIGO}.

\bigskip
 \noindent
 {\it  Big Bang Nucleosynthesis} ~~
The above results indicate that our model is not yet ruled out up to the CMB scale at $z\approx 1100$.   This is still a relatively  low energy scale corresponding to 
a temperature $3000$K.     As stated in the Introduction,   we assume   that 
\eqref{HubT3}  extends to higher $z$,  and this necessarily involves indirect inferences and consequently will be less conclusive.     The next highest energy scale that can provide constraints is
Big Bang Nucleosynthesis (BBN) at $z \approx 10^9$,  corresponding to an energy of about $1$\, MeV and  a temperature of about $10^9$K,      whose consequences can be  inferred from abundances of light nuclei such as Li.   
Based on the formula \eqref{CGdef2},   at this relatively high $z$,   $G_N$ is reduced by  about 40\% if $\bhat =0.02$.    
The constraints on the time variation of $G_N$ for BBN are typically formulated as $\Delta G_N /G_N$,   where $\Delta G_N$ is a bound on how much 
$G_N$ can differ at the time of BBN.    It's important that the reported bounds  assume  slow or no  variation after BBN,    and $\dot{G}_N/G_N$ is based solely on 
a linear fit between the time of BBN and today,  and this should be kept in mind in drawing conclusions.        Incorporating the full time evolution based on \eqref{CGdef2} could significantly alter change these bounds. 
Nevertheless,   let us proceed with some simple checks.     The earliest article is \cite{Krauss}  concludes that to $1\sigma$ confidence level, 
$0.85  <  G_{\rm BBN}/G_N < 1.21$,  which allows a 20\% change of $G_N$ in either direction  in the BBN epoch,  which is borderline consistent with our 
40\% decrease;   as we stated a small change in $\bhat$ could potentially accommodate this.   
To compare with the prediction \eqref{Gdot2},  one should convert   this to a time variation over the intervening time,  and 
  the same article reports  $\dot{G}_N/G_N < 4 \times 10^{-13} \, {\rm yr}^{-1} $ which 
potentially conflicts with  $1.4 \times 10^{-12} \, /{\rm yr}$ based on $H_0$ and $\bhat = 0.02$.     
More recently,     tighter bounds were reported by inferring the consequences of BBN on the CMB \cite{Alvey}. 
With $2\sigma$ confidence level,   this article reports $0.94  <  G_{\rm BBN}/G_N < 1.05$,   which translated to a time evolution one obtains 
$\dot{G}_N/G_N  < 4.5 \times 10^{-12} \, {\rm yr}^{-1}$ \cite{Alvey}.    The latter is consistent with our  $1.4 \times 10^{-12} \, /{\rm yr}$ estimate above.  
We conclude that more work is needed here to determine whether our model is consistent with models of  BBN,   in particular the effect of the complete time evolution which incorporates
the time varying $G_N$ implicit in   \eqref{LCDMHH}.

\subsection{Before Big Bang Nucleosynthesis?} 

At much higher $z$,  experimental probes are more severely  limited and indirect.     Nevertheless some observations 
are worthwhile to point out.   
 As previously discussed,    if the Hubble tension trend extends to this very early universe,   then based on our estimate of $\bhat \approx 0.02$,   
there is a minimal scale factor $\amin$ and a corresponding $\zmax \approx 5 \cdot 10^{21}$,   which corresponds to a hottest temperature 
$\Tmax \approx 10^{22} $K,   which we associated with a hot big bang.    First of all,   at least this $\Tmax$ is above the electro-weak scale of about 
$160$ GeV which is a temperature of about $10^{15}$K,   otherwise the electro-weak transition would have never occurred.   The temperature $\Tmax$ is roughly 
the energy scale of  some Grand Unified Theories,  depending on the model.          In the last section 
we explained how $\zmax$ is roughly the scale for  low-scale  inflation models,   however our model does not exist for scale factors 
$0< a(t) < \amin$,   which is normally considered the inflationary era.   This  led us to suggest  that current models of inflation may not be necessary,   
since at $\zmax$ the Universe is already expanded,   hot,  and flat.    Furthermore,  before the time when $a(t) = \amin$,   our solution to 
$a(t)$ is valid due to  \eqref{minustsymmetry},   namely $a(t) = a(-t + 2 \tmin)$.   A  further constraint on our model is that it should not spoil other positive predictions of inflation,   in particular how quantum density fluctuations of the hypothetical inflaton field can explain the primordial density fluctuations which are thought to provide the seeds for large scale galaxy formation and probed by the CMB.   Quantum fluctuations of  the field for  lightest particle   of mass $\mzeron$,   upon which our formula for 
$\rhovac$ in \eqref{rhovacg} is based upon,  could potentially play such a role,   but such considerations require  further investigation.

\subsection{Bench-top experiments on  the temperature dependence of  Newton's constant}

\def\Tref{\tilde{T}}

 Above we considered implications for  cosmology based on  the $\LambdaCDM$ model and beyond.    Apart from the trend recently observed for
 supernovae discussed above in subsection A,     unfortunately  we could only compare with bounds rather than 
 definitive non-null experimental signatures.        The formula \eqref{CGdef2} implies that Newton's constant actually depends on temperature.  
 An analogy can be made with QED where one must take into account the RG flow of the fine structure constant to make predictions at higher energies.   
 For the $\LambdaCDM$ model,    the energy scales involved are relatively low since $\mu_0 = k_B T_0 = 2.35 \times 10^{-4} \, {\rm eV}$ for $T_0 = 2.7 $\,K.    For cosmology, the temperature  $T$ as described above is the temperature of the entire Universe,    thus  it is not clear if such a 
temperature dependent Newton's constant is also valid locally in space or time.          If it is,   
 and our proposed energy dependent Newton's constant $\CG (\mu)$ in equation 
\eqref{CGdef2}  is correct,  then it is {\it  in principle} possible to
confirm  it in some bench-top Cavendish like experiments as a function of temperature.    If this could be confirmed,  it would be truly remarkable that the new  fundamental parameter $\bhat$ introduced in this article based on the Hubble tension could be measured in a small bench-top experiment.   
Ideally one wishes to measure the gravitational force between two masses as a function of their {\it equal}   temperatures,
however we found no such  such experiments in the literature.       We did find 3 not so well-known experiments wherein the weight of a sample is measured as 
a function of temperature \cite{Shaw,Dmitriev,Liangzao},  the earliest going as far back as 1923,  wherein a positive result is reported rather than a bound.         
Since only the temperature of the sample is varied,  and not that of the Earth,     it's again not clear if our proposal applies directly as stated,  however let us proceed.    
 
Suppose such experiments are carried out in a small range of temperature $\Delta T$ about a fixed 
temperature $\Tref$,   i.e. $T = \Tref + \Delta T$.        From \eqref{CGdef} 
\beq
\label{Cavendish}
\CG (T) = \CG(\Tref) \(1 - \bhat \log (T/\Tref) \)   =   \CG(\Tref) \(1 - \frac{\bhat}{\Tref}  \,  \Delta T  + \frac{\bhat}{2 \Tref^2} \, (\Delta T)^2 + \ldots  \),
\eeq
where $\CG (T)$ is the temperature dependent Newton's constant.   As discussed above,   the effective Newton's constant {\it decreases} with increased temperature in our model since $\bhat >0$.     
If $\Tref = 300^\circ $\,K   then $\bhat/\Tref  \approx 6  \times 10^{-5}$ per degree change 
$\Delta T$ if 
 $\bhat \approx  0.02$, 
where  recall this value of $\bhat$ was inferred from the Hubble tension in \eqref{HubT2}.     
    The experiments  \cite{Shaw,Dmitriev,Liangzao}  indeed consistently  measure a {\it decrease}  in the gravitational force as the temperature is {\it increased},  and 
    all 3 experiments roughly agree in a fit to the $\Delta T$ term in \eqref{Cavendish},    thus our model at the very least predicts the right sign of $\bhat$.            
Based on the ideas in this article,   this decrease should be independent of the material,  in these cases  a metal,   since the Equivalence Principle is built into the formalism,        
whereas experiments find slightly different relative changes in weight  depending on the metal.    
This could be due to experimental uncertainties  as a result of buoyancy,  thermal expansion,  increased mass due to thermal energy,  and other factors which are difficult to take into account.   
There was no theoretical motivation to consider a functional fit based on $1-\bhat \log (T/\Tref)$ in these experiments.      Keeping just the $\Delta T$ term in 
\eqref{Cavendish}  these experiments are roughly consistent  with each other and  yield  variations of about 
$10^{-5}$ per degree change $\Delta T$ for  $\Tref \approx 300$K.       Taking copper for instance \cite{Liangzao},    the ratio of the weights at $T=200^\circ$\,C to that at $T=20^\circ$\,C is
about $0.997$.   Equating this to $1-\bhat \log(473/293)$  gives $\bhat \approx 0.005$,   which is a bit smaller than the value $\bhat \approx 0.02$ based on Hubble tension.
Given  the uncertainties in these rather crude experiments and the value of $\bhat$ itself,   we consider this as positive  support for our model.  
At lower reference temperatures $\Tref$,   the effect is larger and this motivates carrying out such 
 experiments at lower temperatures with more modern experimental techniques.  
     An ideal Cavendish type of experiment as a function of  relatively low temperature could provide a completely independent measurement of the basic parameter $\bhat$ 
to be compared with Hubble tension data.

\def\Tref{\tilde{T}}

\section{Closing Remarks}

Having already summarized the unconventional perspective  on Gravity  developed in this article and its implications in the Introduction,  we close with some  suggestions for further investigations and 
discuss some open questions.     There are  currently many open avenues  for testing (and falsifying) our 
main proposals,  in addition to those considered in the last section.

\bigskip
$\bullet$ ~    The induced RG flow for Newton's constant that we proposed as a consequence of the 
RG flow of the coupling $\gcal$ in $\rhovac$ in \eqref{rhovacg} led to an avoidance of a geometric curvature singularity at a scale factor $a=0$.   There are various theorems asserting  that such singularities are unavoidable in cosmology and black holes \cite{Penrose,HawkingPenrose}.    The latter theorems assume a constant Newton's constant $G_N$,   and should be revisited to account for an effective energy scale dependent $G_N$ as proposed in this article.   This could potentially lead to a resolution of the singularity at the origin of the Schwarzschild black hole solution,  in a way analogous to
the minimal scale factor $\amin$ above.

 \bigskip
 $\bullet$ ~ 
 The minimal scale factor $\amin \sim  10^{-22}$  in eq. \eqref{amin} could produce a
primordial gravitational wave  spectrum detectable by LIGO or eventually the space based interferometer  LISA.     
A weaker  $G_N$ at higher redshift $z$ could delay
stellar collapse, favoring primordial black holes.   In fact,  based on \eqref{GBH},    for a weaker $G_N$,  black holes are smaller for a given mass.      
  Very recently,   supermassive black holes,  with mass on the order of 40 million solar masses have been discovered   using the James Webb Space Telescope \cite{JWST},  which were formed at an earlier time and smaller than expected,    and this may be an interesting topic for further study in the context of the varying $G_N$ that we specifically proposed.

\bigskip
$\bullet$ ~  As stated in the Introduction,  although the formula \eqref{rhovacg} is well-motivated by the works 
\cite{ALrhovac1,ALrhovac2},    we left aside the issue of identifying  the particle with mass $\mzeron$  underlying  this  formula for $\rhovac$ since we were able to 
make progress without doing so,  and the particle physics involved is beyond the original scope of this article.     Nevertheless,  we can constrain some of its properties. 
  We were able to constrain  its mass from the observed value of 
$\rhovac$ in \eqref{mzOfg},    and pointed out that this is consistent with  Majorana neutrinos as viable candidates,   but were not able to make a strong case for this without
a more complete theory of the origin of neutrino masses.   Based on our comparison of coupling constants with QED,   it  could be that the particle with mass
$\mzeron$ is coupled to a hidden U(1) gauge theory where $\gcal$ is a marginally irrelevant fine-structure constant.   
We assumed that our formula for $\rhovac$ in \eqref{rhovacg}  is valid up to $\zmax$,   which would seem to imply that such a particle does not obtain its mass $\mzeron$
from the Higgs mechanism since electro-weak symmetry breaking occurs at  a  significantly lower 
scale of about $160$ GeV,  which corresponds to $z\approx 6 \times 10^{14}$.     In our current understanding of the Standard Model of particle physics, 
 neutrino masses do not arise from the Higgs mechanism since the
Standard Model does not have right-handed neutrinos to pair up with  the known left-handed neutrinos to provide a mass.   
Based on this,  
a  massive Majorana  neutrino is the most promising  candidate for the $\mzeron$-particle,    however other possibilities where this is an entirely new particle are certainly not ruled out. 
   This  underlying QFT involving $\mzeron$ 
  is incomplete in  the UV since it relies on the RG beta function \eqref{betagcal} for a marginally irrelevant coupling,   where the incompleteness is manifested as a Landau pole at very high energies.  
However we showed that the time  evolution of the scale factor $a(t)$ smoothly traverses this Landau  pole through a swing rather than a bang.   
  The UV completeness  issue here  is  certainly more tractable than the UV incompleteness of quantum gravity,    since it is a UV issue of QFT in flat Minkowski space where such problems are much  better understood.     
  
  \bigskip
$\bullet$ ~  In Section VIID we described some not so well-known small scale  bench-top experiments  on the temperature variation of Newton's constant which provided some positive support for our model.      It is potentially exciting that our new parameter $\bhat$ for cosmology can in principle be measured by these completely different kinds of experiments,   and this justifies reproducing these results with more modern experimental techniques.

\section{Acknowledgements}

We wish to  thank Maria Dainotti,  Giovanni Montani, and   Biagio de Simone  of  the NAOJ collaboration for pointing out their recent articles and discussions of their results shortly after the first version of this article appeared.   We also benefitted from  subsequent discussions with Seamus Davis and  Eanna Flanagan.


\begin{thebibliography}{99}

\bibitem{Sakharov}  A. D. Sakharov,
 {\it Vacuum Quantum Fluctuations in Curved Space and the Theory of Gravitation},
 Soviet Physics Doklady, 12, 1040 (1967).

\bibitem{Visser} M. Visser, 
 {\it Sakharov's Induced Gravity: A Modern Perspective"},
 Modern Physics Letters A 17 (2002);  arXiv preprint gr-qc/0204062.   

\bibitem{MisnerThornWheeler}    
C. W. Misner,  K.S. Thorne, and J. A. Wheeler,  {\bf Gravitation,}  Macmillan  (1973)  page 426.



\bibitem{KWilson}  K. G. Wilson, ``Renormalization Group and 
Strong Interactions'', Phys. Rev. {D3} (1971) 1818.

\bibitem{ALrhovac1}   A.  LeClair,   
{\it Thermodynamic formulation of vacuum energy density in flat spacetime and potential implications for the cosmological constant},  JHEP 2024,  arXiv:2404.02350.

\bibitem{ALrhovac2}   A.  LeClair,    
{\it Vacuum energy density from the form factor bootstrap},   JHEP 2024,   arXiv:2407.10692.


\bibitem{Planck}  
N. Aghanim et al.,
{\it Planck 2018 results. VI. Cosmological parameters,}   Astronomy and  Astrophysics, 641, A6. (2020).

\bibitem{NeutrinoMasses}  M. C. Gonzalez-Garcia and Y. Nir, Neutrino masses and mixing: evidence and implications, Rev. Mod. Phys. 75 (2003)
345, arXiv:hep-ph/0202058. 



\bibitem{Bekenstein}
J. D. Bekenstein,
{\it  Black Holes and Entropy,}
 Physical Review D 7, 2333–2346 (1973).

\bibitem{Hawking}
S. W. Hawking, 
{\it Particle creation by black holes,}
 Communications in Mathematical Physics, Volume 43, 199–220 (1975).


\bibitem{GibbonsHawking} 
G. W. Gibbons and S. W. Hawking, 
{\it  Cosmological Event Horizons, Thermodynamics, and Particle Creation,} Physical Review D, 15(10), 2738–2751 (1977).


\bibitem{StringTheory}  
M. B. Green, J. H. Schwarz, and E. Witten. {\it Superstring theory}, Cambridge university press (2012). 

\bibitem{Verlinde}  E. Verlinde,
{\it On the origin of gravity and the laws of Newton}, 
JHEP 2011(4),  arXiv:1001.0785 .


\bibitem{Feynman}   
D. Rickles and C. M.  DeWitt  (eds.).
{\it  The Role of Gravitation in Physics: Report from the 1957 Chapel Hill Conference. Edition Open Access, 2011,}   (Available online at: https://edition-open-sources.org/sources/5/).  


 \bibitem{Strominger}     D. Anninos,   T.  Hartman and A.  Strominger,   
 {\it Higher Spin Realization of the dS/CFT Correspondence,}
 Class. Quant. Grav. {\bf 34} (2016) 015009;   arXiv:1108.5735.


\bibitem{Oppenheim}   
J.  Oppenheim, 
{\it A Postquantum Theory of Classical Gravity?},  Phys. Rev. X 13, 041040 (2023);
arXiv:1811.03116.  



\bibitem{LeClairUniverse}  A. LeClair,
{\it  Comment on the cosmological constant for $\lambda \phi^4$ theory in $D$ spacetime dimensions,}
Universe (2023) 310  arXiv:2304.13075 [hep-th].







\bibitem{Casimir}
H. B. G. Casimir, 
{\it On the attraction between two perfectly conducting plates,}   Proceedings of the Koninklijke Nederlandse Akademie van Wetenschappen, 51, 793–795 (1948).    


\bibitem{GibbonsDdim} 
S. Chen, G. W. Gibbons, Y. Li and Y. Yang,
{\it  Friedmann’s equations in all dimensions and Chebyshev’s theorem,}  Journal
of Cosmology and Astroparticle Physics (2014) 035.

\bibitem{JTgravity}   
R. Jackiw and C. Teitelboim, 
{\it Two-dimensional gravity and nonlinear gauge theory,}
 In Quantum Theory of Gravity: Essays in Honor of the 60th Birthday of Bryce S. DeWitt (pp. 389–398). Adam Hilger.
(1985).




\bibitem{Weinberg}
S. Weinberg, 
{\it The Cosmological Constant Problem,}  Rev. Mod. Phys. 61 (1989) 1.


\bibitem{ZamoZamo0}  A. B. Zamolodchikov and A. B. Zamolodchikov, 
{\it Factorized S-matrices in two dimensions as the exact solutions of certain relativistic quantum field theory models}, 
Annals of Physics 120 (1979)  273.  

\bibitem{YangYang}   C. N. Yang and C. P. Yang,  
{\it Thermodynamics of a one-dimensional system of bosons with repulsive delta-function interaction,}
J. Math. Phys. 10 (1969) 1115.   

\bibitem{ZamoTBA}  
Al. B.  Zamolodchikov, {\it Thermodynamic Bethe Ansatz in relativistic models: scaling 3-state Potts nd Lee-Yang models}, 
Nucl.  Phys.  B342 (1990) 695.

\bibitem{KlassenMelzer}   
T. Klassen and E. Melzer, {\it The thermodynamics of purely elastic scattering theories and conformal perturbation theory,} 
Nucl. Phys. B350 (1991) 635.

\bibitem{Corrigan}   H. W. Braden, E. Corrigan, P. E. Dorey, and R. Sasaki, 
{\it Aﬃne Toda field theory and exact S-matrices,}  Nucl. Phys. B338
(1990) 689.

\bibitem{DestrideVega} 
C. Destri and H. J. de Vega,
{\it New exact results in Affine Toda field theories: Free energy and wave- function renormalizations}, Nuclear Physics B358 (1991) 251-294.


\bibitem{ZamoShG} Al. B. Zamolodchikov, 
{\it Mass scale in the sine-Gordon model and its reductions,}  Int. J. Mod. Phys. A10 (1995) 1125.

\bibitem{Peskin}
M. E. Peskin and D. V. Schroeder, An Introduction to Quantum Field Theory, Addison-Wesley 1995.


\bibitem{Festina1}
M. Montero, T. Van Riet and G. Venken, 
{\it Festina Lente: EFT Constraints from Charged Black Hole Evaporation,}  JHEP
2020.1: 1-50, arXiv:1910.01648 [hep-th].



\bibitem{Festina2} 
M. Montero, C. Vafa, T. Van Riet and G. Venken, 
{\it The FL bound and its phenomenological implications,}  JHEP 10 (2021)
009, arXiv:2106.07650 [hep-th]


\bibitem{LandauPole1} 
M. G\"ockeler, R. Horsley, V.  Linke, P. Rakow,  G. Schierholz and H. St\"uben,
{\it  Is there a Landau pole problem in QED?},  Physical Review Letters  80, 4119 (1998). 

 
 \bibitem{LandauPole2}
H.  Gies and J.  Jaeckel, 
{\it Renormalization flow of QED,}  Physical Review Letters 93  (2004) 110405.

\bibitem{LandauPole3}
S.-K. Jian,  E.  Barnes, and S. D.  Sarma,
{\it Landau poles in condensed matter systems,} Physical Review Research 2.2 (2020): 023310.


\bibitem{SHOES} 
A. G. Riess  et al,
{\it  A Comprehensive Measurement of the Local Value of the Hubble Constant with $1\% $  Precision from the Hubble Space Telescope and the SH0ES Team,}  The Astrophysical Journal Letters, 934(1)  (2022).   

\bibitem{Riess2}
L. Verde,  T.  Treu and A. G. Riess,
{\it  Tensions between the early and late Universe,}
 Nature Astronomy 3.10 (2019): 891-895. 

\bibitem{HubbleTensionRef} 
L. Verde,  T.  Treu,  and A. G.  Riess,  
{\it  Tensions in cosmology: A Nobel perspective on the Hubble constant,}  Nature Reviews Physics, 5, 391–404. (2023).  

 \bibitem{Dainotti2021}   M. G. Dainotti,  et al.
  {\it On the Hubble constant tension in the SNe Ia Pantheon sample,} 
  The Astrophysical Journal 912.2 (2021): 150,   arXiv:2103.02117.

\bibitem{Dainotti2022}
M. G. Dainotti,  et al. 
{\it On the evolution of the Hubble constant with the SNe Ia Pantheon sample and baryon acoustic oscillations: a feasibility study for GRB-cosmology in 2030,}
 Galaxies 10.1 (2022): 24,  arXiv:2201.09848.
 
 \bibitem{Dainotti2024}
B.  De Simone,  et al. 
{\it A doublet of cosmological models to challenge the H0 tension in the Pantheon Supernovae Ia catalog,} 
Journal of High Energy Astrophysics 45 (2025): 290-298,   arXiv:2411.05744. 

 
\bibitem{Dainotti2025}
M. G. Dainotti,   et al. 
{\it A New Master Supernovae Ia sample and the investigation of the $ H_0 $ tension,} 
Journal of High Energy Astrophysics (2025),  arXiv:2501.11772.
  
\bibitem{Montani} G. Montani, E. Fazzari,  N.  Carlevaro anf M. G. Dainotti,
{\it Two Dynamical Scenarios for Binned Master
Sample Interpretation},   Entropy 27.9 (2025) 895.  
  
\bibitem{Montani2}  E. Fazzari,  M. G. Dainotti,  G. Montani and A.  Melchiorri, 
{\it The effective running Hubble constant in SNe Ia as a marker for the dark energy nature},
Journal of High Energy Astrophysics, Vol.  49 (2026),    arXiv:2506.04162.   

\bibitem{ALMaria}   M. Dainotti,  A. Banerjee,  A. LeClair and G. Montani,    
{\it Logarithmic vs Power-Law Expansion from the Binned Master Sample of SNe Ia,}  arXiv preprint arXiv:2603.00497 (2026).





\bibitem{Taylor}
J. M. Weisberg and J. H. Taylor, 
{\it The Relativistic Binary Pulsar B1913+16: Thirty Years of Observations and Analysis,}, ASP Conference Series, vol. 328, pp. 25–31, 2005.

\bibitem{LIGO} B. P. Abbott et al. (LIGO Scientific Collaboration and Virgo Collaboration), 
{\it Tests of General Relativity with GW150914,} 
 Physical Review Letters, vol. 116, no. 22, 221101 (2016).   




\bibitem{Krauss}
C. J. Copi, A. N. Davis, and L. M. Krauss, 
{\it New nucleosynthesis constraint on the variation of G,}   Physical Review Letters 92.17 (2004): 171301;   arXiv:astro-ph/0311334.



\bibitem{Alvey} 
J. Alvey, N. Sabti, M.  Escudero and M Fairbairn,
{\it Improved BBN constraints on the variation of the gravitational constant,}   The European Physical Journal C, 80 (2020);
arXiv:1910.10730. 


\bibitem{Shaw}
P. E. Shaw and N. Davy, 
{\it  The Effect of Temperature on Gravitative Attraction,} 
 Physical Review, 21(6), 680–681  (1923). 



\bibitem{Dmitriev} 
A. L. Dmitriev,  E. M.  Nikushchenko,  and V.S. Snegov, 
{\it  Influence of the Temperature of a Body on Its Weight,}  Measurement Techniques, 46(2), 115–120 (2003).   

\bibitem{Liangzao}
F. Liangzao, F.  Jinsong, and  L. W.  Qing,
{\it An Experimental Discovery about Gravitational Force Changes in Materials due to Temperature Variation,}   Engineering Sciences, 8(2), 9–11. (2010).  


\bibitem{Penrose} 
R. Penrose, 
{\it Gravitational Collapse and Space-Time Singularities.} Physical Review Letters, 14(3), 57–59 (1965).

\bibitem{HawkingPenrose}  
S. W. Hawking and R. Penrose,
{\it The Singularities of Gravitational Collapse and Cosmology,}  Proceedings of the Royal Society of London. Series A, Mathematical and Physical Sciences, 314(1519), 529–548 (1970).



\bibitem{JWST} 
R. Maiolino  et al. 
{\it A small and vigorous black hole in the early Universe,} 
 Nature, 627, 70059 (2024). 

\bibitem{JHEAP}  A.  LeClair, 
{\it Quantum Vacuum energy as the origin of Gravity},   Journal of High Energy Astrophysics (2026): 100546.

\end{thebibliography}
\end{document}